\documentclass[11pt]{article}%
\usepackage{makeidx}
\usepackage{graphicx}
\usepackage{multicol}
\usepackage[bottom]{footmisc}
\usepackage{latexsym}
\usepackage{subfigure}
\usepackage{cite,color,comment,xspace}
\usepackage{amsfonts}
\usepackage{amsmath}
\usepackage{amssymb}
\usepackage{algorithm}
\usepackage{algorithmic}
\usepackage{url}
\usepackage{times}
\usepackage{mathptmx}
\usepackage{enumerate}
\usepackage{epstopdf}
\usepackage{subfigure}
\usepackage{epsfig}%
\providecommand{\U}[1]{\protect\rule{.1in}{.1in}}
\newtheorem{theorem}{Theorem}

\newtheorem{proposition}[theorem]{Proposition}

\makeatletter

\@addtoreset{theorem}{section}

\@addtoreset{corollary}{section}

\@addtoreset{lemma}{section}

\@addtoreset{definition}{section}
\makeatother
\begin{document}

\title{{\LARGE \textbf{An Optimal Control Approach to the Persistent Monitoring
Problem\\ - Technical Report - }}}
\author{\textbf{Christos.G. Cassandras, Xuchao Lin }and\textbf{ Xu Chu Ding}%
\thanks{{\footnotesize The authors' work is supported in part by NSF under
Grant EFRI-0735974, by AFOSR under grant FA9550-09-1-0095, by DOE under grant
DE-FG52-06NA27490, by ONR under grant N00014-09-1-1051 and by ARO under grant
W911NF-11-1-0227.} }\\Division of Systems Engineering\\and Center for Information and Systems Engineering\\Boston University, cgc@bu.edu, mmxclin@bu.edu, xcding@bu.edu}
\date{January 2012}
\maketitle

\begin{abstract}
We propose an optimal control framework for persistent
monitoring problems where the objective is to control the movement of mobile
nodes to minimize an uncertainty metric in a given mission space. For multi
agent in a one-dimensional mission space, we show that the optimal solution is
obtained in terms of a sequence of switching locations and waiting time on
these switching points, thus reducing it to a parametric optimization problem.
Using Infinitesimal Perturbation Analysis (IPA) we obtain a complete solution
through a gradient-based algorithm. We also discuss a receding horizon
controller which is capable of obtaining a near-optimal solution on-the-fly.

\end{abstract}

%\title{Using Infinitesimal Perturbation Analysis of Stochastic Flow Models to Recover
%Performance Sensitivity Estimates of Discrete Event Systems}
%\author{\textbf{Chen Yao }and\textbf{ Christos G. Cassandras}%
%\thanks{{\footnotesize Supported in part by by the National Science Foundation
%under Grant EFRI-0735794, by AFOSR under grants FA9550-07-1-0361 and
%FA9550-09-1-0095, by DOE under grant DE-FG52-06NA27490, and by ONR under grant
%N00014-09-1-1051.}}\\Division of Systems Engineering\\and Center for Information and Systems Engineering\\[0pt] Boston University\\[0pt] Brookline, MA 02446\\[0pt] \texttt{cyao@bu.edu, cgc@bu.edu}}
%\maketitle

\section{Introduction}

Enabled by recent technological advances, the deployment of autonomous agents
that can cooperatively perform complex tasks is rapidly becoming a reality. In
particular, there has been considerable progress reported in the literature on
robotics and sensor networks regarding coverage control
\cite{rekleitis2004limited,cortes2004coverage,li2006cooperative}, surveillance
\cite{girard2005border,grocholsky2006cooperative} and environmental sampling
\cite{smith2012persistent,paley2008cooperative} missions. In this paper, we
are interested in generating optimal control strategies for \emph{persistent
monitoring} tasks; these arise when agents must monitor a dynamically changing
environment which cannot be fully covered by a stationary team of available
agents. Persistent monitoring differs from traditional coverage tasks due to
the perpetual need to cover a changing environment, i.e., all areas of the
mission space must be visited infinitely often. The main challenge in
designing control strategies in this case is in balancing the presence of
agents in the changing environment so that it is covered over time optimally
(in some well-defined sense) while still satisfying sensing and motion
constraints. Examples of persistent monitoring missions include surveillance
and theft prevention in a building, patrol missions with unmanned vehicles,
and environmental applications where routine sampling of an area is involved.

In this paper, we address the persistent monitoring problem by proposing an
optimal control framework to drive agents so as to minimize a metric of
uncertainty over the environment. In coverage control
\cite{cortes2004coverage,li2006cooperative}, it is common to model knowledge
of the environment as a non-negative density function defined over the mission
space, and usually assumed to be fixed over time. However, since persistent
monitoring tasks involve dynamically changing environments, it is natural to
extend it to a function of both space and time to model uncertainty in the
environment. We assume that uncertainty at a point grows in time if it is not
covered by any agent sensors. To model sensor coverage, we define a
probability of detecting events at each point of the mission space by agent
sensors. Thus, the uncertainty of the environment decreases with a rate
proportional to the event detection probability, i.e.,\textit{ } the higher
the sensing effectiveness is, the faster the uncertainty is reduced..

While it is desirable to track the value of uncertainty over all points in the
environment, this is generally infeasible due to computational complexity and
memory constraints. Motivated by polling models in queueing theory, e.g.,
spatial queueing \cite{bertsimas1993stochastic},\cite{cooper1981introduction},
and by stochastic flow models \cite{sun2004perturbation}, we assign sampling
points of the environment to be monitored persistently (this is equivalent to
partitioning the environment into a discrete set of regions.) We associate to
these points \textquotedblleft uncertainty queues\textquotedblright\ which are
visited by one or more \textquotedblleft servers\textquotedblright.\ The
growth in uncertainty at a sampling point can then be viewed as a flow into a
queue, and the reduction in uncertainty (when covered by an agent) can be
viewed as the queue being visited by mobile servers as in a polling system.
Moreover, the service flow rates depend on the distance of the sampling point
to nearby agents. From this point of view, we aim to control the movement of
the servers (agents) so that the total accumulated \textquotedblleft
uncertainty queue\textquotedblright\ content is minimized.

Control and motion planning for agents performing persistent monitoring tasks
have been studied in the literature. In \cite{rekleitis2004limited} the focus
is on sweep coverage problems, where agents are controlled to sweep an area.
In \cite{smith2012persistent,nigam2008persistent} a similar metric of
uncertainty is used to model knowledge of a dynamic environment. In
\cite{nigam2008persistent}, the sampling points in a 1-dimensional environment
are denoted as cells, and the optimal control policy for a two-cell problem is
given. Problems with more than two cells are addressed by a heuristic policy.
In \cite{smith2012persistent}, the authors proposed a stabilizing speed
controller for a single agent so that the accumulated uncertainty over a given
path in the environment is bounded, along with an optimal controller that
minimizes the maximum steady-state uncertainty, assuming that the agent
travels along a closed path and does not change direction. The persistent
monitoring problem is also related to robot patrol problems, where a team of
robots are required to visit points in the workspace with frequency
constraints
\cite{hokayem2008persistent,elmaliach2008realistic,elmaliach2007multi}.

Our ultimate goal is to optimally control a team of cooperating agents in a 2
or 3-dimensional environment. The contribution of this paper is to take a
first step toward this goal by formulating and solving an optimal control
problem for a team of agents moving in a 1-dimensional mission space described
by an interval $[0,L]\subset\mathbb{R}$ in which we minimize the accumulated
uncertainty over a given time horizon and over an arbitrary number of sampling
points. Even in this simple case, determining a complete explicit solution is
computationally hard. However, we show that the problem can be reduced to a
\emph{parametric} optimization problem. In particular, the optimal trajectory
of each agent is to move at full speed until it reaches some switching point,
dwell on the switching point for some time (possibly zero), and then switch
directions. In addition, we prove that all agents should never reach the end
points of the mission space $[0,L]$. Thus, each agent's optimal trajectory is
fully described by a set of switching points $\{\theta_{1},\ldots,\theta
_{K}\}$ and associated waiting times at these points, $\{w_{1},\ldots,w_{K}%
\}$. As a result, we show that the behavior of the agents operating under
optimal control is described by a hybrid system. This allows us to make use of
generalized Infinitesimal Perturbation Analysis (IPA), as presented in
\cite{cassandras2009perturbation},\cite{Wardietal09},\textbf{ }to determine
gradients of the objective function with respect to these parameters and
subsequently obtain optimal switching locations and waiting times that fully
characterize an optimal solution. It also allows us to exploit robustness
properties of IPA to extend this solution approach to a stochastic uncertainty
model. Our analysis establishes the basis for extending this approach to a
2-dimensional mission space (in ongoing research).\textbf{ }In a broader
context, our approach brings together optimal control, hybrid systems, and
perturbation analysis techniques in solving a class of problems which, under
optimal control, can be shown to behave like hybrid systems characterized by a
set of parameters whose optimal values deliver a complete optimal control solution.

%We then apply a receding horizon approach.
%In this paper, to simplify analysis, we consider setting with one node  Our controller, using optimal control method,
%guarantee to converge to local (in terms of switching locations positions)
%minimum uncertainty value for the whole workspace. We first prove that the
%optimal control node will always moves with maximum speed, and then we show
%that the optimal node will never hit the two end points. These two conclusions
%generalize that the task of designing the optimal controller is to find the
%optimal switching locations for any given operation time.
%The contributions of this paper are to lay out a formal formulation for 1-D
%persistent monitoring problem, prove the properties that one node optimal
%controller should have, and use IPA to find the optimal switching locations. Also
%our experiment results have shown that for longer operation time, searching
%optimal switching locations using gradient descent method can only guarantee to
%converge to local minimum uncertainy value, but not the global minimum.
%Different initial switching locations may end up with different final switching
%points. They are all optimal in terms of local minimum. But since gradient
%descent searching surface is not convex, we have to live with these local minimum.

The rest of the paper is organized as follows. Section
\ref{sec:problemformulation} formulates the optimal control problem. Section
\ref{sec:optimalsolution} characterizes the solution of the optimal control
problem in terms of two parameter vectors specifying switching points in the
mission space and associated dwelling times at them. Using IPA in conjunction
with a gradient-based algorithm, a complete solution is also provided. Section
\ref{sec:exper} provides some numerical results and Section \ref{sec:concl}
concludes the paper.

\section{Persistent Monitoring Problem Formulation}

\label{sec:problemformulation} We consider $N$ mobile agents moving in a
1-dimensional mission space of length $L$, for simplicity taken to be an
interval $[0,L]\subset\mathbb{R}$. Let the position of the agents at time $t$
be $s_{n}(t)\in\left[  0,L\right]  $, $n=1,\ldots,N$, following the dynamics:%
\begin{equation}
\dot{s}_{n}(t)=u_{n}(t) \label{eq:multiDynOfS}%
\end{equation}
i.e., we assume that the agent can control its direction and speed. Without
loss of generality, after some rescaling with the size of the mission space
$L$, we further assume that the speed is constrained by $\left\vert
u_{n}\left(  t\right)  \right\vert \leq1$, $n=1,\ldots,N$. For the sake of
generality, we include the additional constraint:%
\begin{equation}
a\leq s(t)\leq b\text{, \ }a\geq0\text{, \ }b\leq L \label{NoEndpoints}%
\end{equation}
over all $t$ to allow for mission spaces where the agents may not reach the
end points of $\left[  0,L\right]  $, possibly due to the presence of
obstacles. We also point out that the agent dynamics in (\ref{eq:multiDynOfS})
can be replaced by a more general model of the form $\dot{s}_{n}%
(t)=g_{n}(s_{n})+b_{n}u_{n}(t)$ without affecting the main results of our
analysis (see also Remark 1in Section \ref{sec:Hamiltonian}.) Finally, an
additional constraint may be imposed if we assume that the agents are
initially located so that $s_{n}\left(  0\right)  <s_{n+1}\left(  0\right)  $,
$n=1,\ldots,N-1$, and we wish to prevent them from subsequently crossing each
other over all $t$:%
\begin{equation}
s_{n}\left(  t\right)  -s_{n+1}\left(  t\right)  \leq0 \label{eq:multiNoCross}%
\end{equation}
We associate with every point $x\in\left[  0,L\right]  $ a function
$p_{n}(x,s_{n})$ that measures the probability that an event at location $x$
is detected by agent $n$. We also assume that $p_{n}(x,s_{n})=1$ if $x=s_{n}$,
and that $p_{n}(x,s_{n})$ is monotonically nonincreasing in the distance
$|x-s_{n}|$ between $x$ and $s_{n}$, thus capturing the reduced effectiveness
of a sensor over its range which we consider to be finite and denoted by
$r_{n}$ (this is the same as the concept of \textquotedblleft sensor
footprint\textquotedblright\ found in the robotics literature.) Therefore, we
set $p_{n}(x,s_{n})=0$ when $|x-s_{n}|>r_{n}$. Although our analysis is not
affected by the precise sensing model $p_{n}(x,s_{n})$, we will limit
ourselves to a linear decay model as follows:%
\begin{equation}
p_{n}(x,s_{n})=\left\{
\begin{array}
[c]{cc}%
1-\frac{\left\vert x-s_{n}\right\vert }{r_{n}}, & \text{if }|x-s_{n}|\text{
}\leq r_{n}\\
0, & \text{if }|x-s_{n}|\text{ }>r_{n}%
\end{array}
\right.  \label{eq:multiLinearModel}%
\end{equation}
Next, consider a set of points $\{\alpha_{i}\}$, $i=1,\ldots,M$, $\alpha
_{i}\in\lbrack0,L]$, and associate a time-varying measure of uncertainty with
each point $\alpha_{i}$, which we denote by $R_{i}(t)$. Without loss of
generality, we assume $0\leq\alpha_{1}\leq\cdots\leq\alpha_{M}\leq L$ and, to
simplify notation, we set $p_{n,i}(s_{n}(t))\equiv p_{n}(\alpha_{i}%
,s_{n}(t)).$ This set may be selected to contain points of interest in the
environment, or sampled points from the mission space. Alternatively, we may
consider a partition of $[0,L]$ into $M$ intervals whose center points are
$\alpha_{i}=\frac{(2i-1)L}{2M}$, $i=1,\ldots,M$. We can then set
$p_{n}(x,s_{n}\left(  t\right)  )=p_{n,i}(s_{n}\left(  t\right)  )$ for all
$x\in\lbrack\alpha_{i}-\frac{L}{2M},\alpha_{i}+\frac{L}{2M}]$. Therefore, the
joint probability of detecting an event at location $x\in\lbrack\alpha
_{i}-\frac{L}{2M},\alpha_{i}+\frac{L}{2M}]$ by all the $N$ agents
simultaneously (assuming detection independence) is:%
\begin{equation}
P_{i}\left(  \mathbf{s}(t)\right)  =1-%
%TCIMACRO{\dprod \limits_{n=1}^{Q}}%
%BeginExpansion
{\displaystyle\prod\limits_{n=1}^{Q}}
%EndExpansion
\left[  1-p_{n,i}(s_{n}\left(  t\right)  )\right]  \label{DefOfPro}%
\end{equation}
where we set $\mathbf{s}(t)=[s_{1}\left(  t\right)  ,\ldots,s_{N}\left(
t\right)  ]^{\text{T}}$. We define uncertainty functions $R_{i}(t)$ associated
with the intervals $[\alpha_{i}-\frac{L}{2M},\alpha_{i}+\frac{L}{2M}]$,
$i=1,\ldots,M$, so that they have the following properties: $(i)$ $R_{i}(t)$
increases with a prespecified rate $A_{i}$ if $P_{i}\left(  \mathbf{s}%
(t)\right)  =0$, $(ii)$ $R_{i}(t)$ decreases with a fixed rate $B$ if
$P_{i}\left(  \mathbf{s}(t)\right)  =1$ and $(iii)$ $R_{i}(t)\geq0$ for all
$t$. It is then natural to model uncertainty so that its decrease is
proportional to the probability of detection. In particular, we model the
dynamics of $R_{i}(t)$, $i=1,\ldots,M$, as follows:
\begin{equation}
\dot{R}_{i}(t)=\left\{
\begin{array}
[c]{ll}%
0 & \text{if }R_{i}(t)=0,\text{ }A_{i}\leq BP_{i}\left(  \mathbf{s}(t)\right)
\\
A_{i}-BP_{i}\left(  \mathbf{s}(t)\right)  & \text{otherwise}%
\end{array}
\right.  \label{eq:multiDynR}%
\end{equation}
where we assume that initial conditions $R_{i}(0)$, $i=1,\ldots,M$, are given
and that $B>A_{i}>0$ (thus, the uncertainty strictly decreases when there is
perfect sensing $P_{i}\left(  \mathbf{s}(t)\right)  =1$.)

\begin{figure}[ptb]%
\centering
\includegraphics[
natheight=3.195500in,
natwidth=6.005300in,
height=2.0781in,
width=3.8869in
]%
{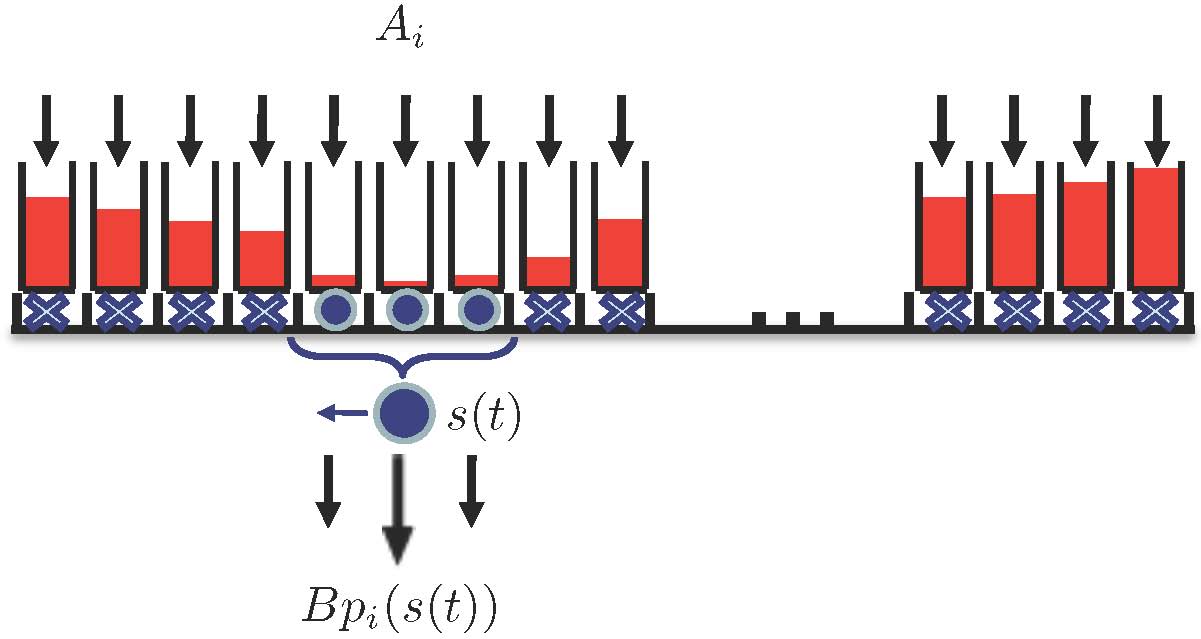}%
\caption{A queueing system analog of the persistent monitoring problem.}%
\label{fig:queue}%
\end{figure}

Viewing persistent monitoring as a polling system, each point $\alpha_{i}$
(equivalently, $i$th interval in $[0,L]$) is associated with a
\textquotedblleft virtual queue\textquotedblright\ where uncertainty
accumulates with inflow rate $A_{i}$. The service rate of this queue is
time-varying and given by $BP_{i}\left(  \mathbf{s}(t)\right)  $, controllable
through the agent position at time $t$. Figure \ref{fig:queue} illustrates
this polling system when $N=1$. This interpretation is convenient for
characterizing the \emph{stability} of such a system over a mission time $T$:
For each queue, we may require that $\int_{0}^{T}A_{i}<\int_{0}^{T}%
Bp_{i}(s(t))dt$. Alternatively, we may require that each queue becomes empty
at least once over $[0,T]$. We may also impose conditions such as
$R_{i}(T)\leq R_{\max}$ for each queue as additional constraints for our
problem so as to provide bounded uncertainty guarantees, although we will not
do so in this paper. Note that this analogy readily extends to 2 or
3-dimensional settings.%

%TCIMACRO{\FRAME{ftbpFU}{3.8869in}{2.0781in}{0pt}{\Qcb{A queueing system analog
%of the persistent monitoring problem.}}{\Qlb{fig:queue}}{queue.jpg}%
%{\special{ language "Scientific Word";  type "GRAPHIC";
%maintain-aspect-ratio TRUE;  display "USEDEF";  valid_file "F";
%width 3.8869in;  height 2.0781in;  depth 0pt;  original-width 6.0053in;
%original-height 3.1955in;  cropleft "0";  croptop "1";  cropright "1";
%cropbottom "0";  filename '../journal_v5/queue.jpg';file-properties "XNPEU";}%
%}}%
%BeginExpansion
%EndExpansion

The goal of the optimal persistent monitoring problem we consider is to
control the movement of the $N$ agents through $u_{n}\left(  t\right)  $ in
(\ref{eq:multiDynOfS}) so that the cumulative uncertainty over all sensing
points $\{\alpha_{i}\},$ $i=1,\ldots,M$ is minimized over a fixed time horizon
$T$. Thus, setting $\mathbf{u}\left(  t\right)  =\left[  u_{1}\left(
t\right)  ,\ldots,u_{N}\left(  t\right)  \right]  $ we aim to solve the
following optimal control problem\textbf{ P1}:
\begin{equation}
\min_{\mathbf{u}\left(  t\right)  }\text{ \ }J=\frac{1}{T}\int_{0}^{T}%
\sum_{i=1}^{M}R_{i}(t)dt\label{eq:costfunction}%
\end{equation}
subject to the agent dynamics \eqref{eq:multiDynOfS}, uncertainty dynamics
\eqref{eq:multiDynR}, control constraint $|u_{n}(t)|\leq1$, $t\in\lbrack0,T]$,
and state constraints (\ref{NoEndpoints}), $t\in\lbrack0,T]$. Note that we
require $a\leq r_{n}$ and $b\geq L-r_{m}$, for at least some $n,m=1,\ldots,N$;
this is to ensure that there are no points in $[0,L]$ which can never be
sensed, i.e., any $i$ such that $\alpha_{i}<a-r_{n}$ or $\alpha_{i}>b+r_{n}$
would always lie outside any agent's sensing range. We will omit the
additional constraint (\ref{eq:multiNoCross}) from our initial analysis, but
we will show that, when it is included, the optimal solution never allows it
to be active.

\section{Optimal Control Solution}

\label{sec:optimalsolution}

\subsection{Hamiltonian analysis}

\label{sec:Hamiltonian}

We first characterize the optimal control solution of problem\textbf{ P1} and
show that it can be reduced to a parametric optimization problem. This allows
us to utilize an Infinitesimal Perturbation Analysis (IPA) gradient estimation
approach \cite{cassandras2009perturbation} to find a complete optimal solution
through a gradient-based algorithm. We define the state vector $\mathbf{x}%
\left(  t\right)  =[s_{1}\left(  t\right)  ,\ldots,s_{N}\left(  t\right)
,R_{1}\left(  t\right)  ,\ldots,R_{M}\left(  t\right)  ]^{\mathtt{T}}$ and the
associated costate vector $\mathbf{\lambda}\left(  t\right)  =$ $[\lambda
_{s_{1}}\left(  t\right)  ,\ldots,\lambda_{s_{N}}\left(  t\right)
,\lambda_{1}\left(  t\right)  ,\ldots,\lambda_{M}\left(  t\right)
]^{\mathtt{T}}$. In view of the discontinuity in the dynamics of $R_{i}(t)$ in
(\ref{eq:multiDynR}), the optimal state trajectory may contain a boundary arc
when $R_{i}(t)=0$ for any $i$; otherwise, the state evolves in an interior
arc. We first analyze the system operating in such an interior arc and omit
the constraint (\ref{NoEndpoints}) as well. Using (\ref{eq:multiDynOfS}) and
(\ref{eq:multiDynR}), the Hamiltonian is%
\begin{equation}
H\left(  \mathbf{x},\mathbf{\lambda},\mathbf{u}\right)  =\sum_{i=1}^{M}%
R_{i}\left(  t\right)  +\sum_{n=1}^{N}\lambda_{s_{n}}\left(  t\right)
u_{n}\left(  t\right)  +\sum_{i=1}^{M}\lambda_{i}\left(  t\right)  \dot{R}%
_{i}(t) \label{eq:Hamiltonian}%
\end{equation}
and the costate equations $\mathbf{\dot{\lambda}}=-\frac{\partial H}{\partial
x}$ are%
\begin{align}
\dot{\lambda}_{i}\left(  t\right)   &  =-\frac{\partial H}{\partial
R_{i}\left(  t\right)  }=-1\text{, \ \ }i=1,\ldots,M\label{eq:multiDynCoI}\\
\dot{\lambda}_{s_{n}}\left(  t\right)   &  =-\frac{\partial H}{\partial
s_{n}\left(  t\right)  }=-\frac{B}{r_{n}}\sum_{i\in\digamma_{n}^{-}\left(
t\right)  }\lambda_{i}\left(  t\right)
%TCIMACRO{\dprod \limits_{d\neq n}}%
%BeginExpansion
{\displaystyle\prod\limits_{d\neq n}}
%EndExpansion
\left[  1-p_{d,i}(s_{d}\left(  t\right)  )\right]  +\frac{B}{r_{n}}\sum
_{i\in\digamma_{n}^{+}\left(  t\right)  }\lambda_{i}\left(  t\right)
%TCIMACRO{\dprod \limits_{d\neq n}}%
%BeginExpansion
{\displaystyle\prod\limits_{d\neq n}}
%EndExpansion
\left[  1-p_{d,i}(s_{d}\left(  t\right)  )\right]  \label{eq:multiDynCoS}%
\end{align}
where we have used (\ref{eq:multiLinearModel}), and the sets $\digamma_{n}%
^{-}(t)$ and $\digamma_{n}^{+}(t)$ are defined as%
\begin{align}
\digamma_{n}^{-}(t)  &  =\{i:s_{n}\left(  t\right)  -r_{n}\leq\alpha_{i}\leq
s_{n}\left(  t\right)  \}\text{ \ }\label{eq:multiFset}\\
\digamma_{n}^{+}(t)  &  =\{i:s_{n}\left(  t\right)  <\alpha_{i}\leq
s_{n}\left(  t\right)  +r_{n}\}\nonumber
\end{align}
for $n=1,\ldots,N$. Note that $\digamma_{n}^{-}(t)$, $\digamma_{n}^{+}(t)$
identify all points $\alpha_{i}$ to the left and right of $s_{n}\left(
t\right)  $ respectively that are within agent $n$'s sensing range. Since we
impose no terminal state constraints, the boundary conditions are $\lambda
_{i}\left(  T\right)  =0$, \ $i=1,\ldots,M$ and $\lambda_{s_{n}}\left(
T\right)  =0$, \ $n=1,...,N.$ Applying the Pontryagin minimum principle to
(\ref{eq:Hamiltonian}) with $\mathbf{u}^{\star}(t)$, $t\in\lbrack0,T)$,
denoting an optimal control, we have%
\[
H\left(  \mathbf{x}^{\star},\mathbf{\lambda}^{\star},\mathbf{u}^{\star
}\right)  =\min_{u_{n}\in\lbrack-1,1],\text{ }n=1,\ldots,N}H\left(
\mathbf{x},\mathbf{\lambda},\mathbf{u}\right)
\]
and it is immediately obvious that it is necessary for an optimal control to
satisfy:
\begin{equation}
u_{n}^{\star}(t)=\left\{
\begin{array}
[c]{ll}%
1 & \text{ if }\lambda_{s_{n}}\left(  t\right)  <0\\
-1 & \text{ if }\lambda_{s_{n}}\left(  t\right)  >0
\end{array}
\right.  \label{eq:multiUstar}%
\end{equation}
This condition excludes the possibility that $\lambda_{s_{n}}\left(  t\right)
=0$ over some finite singular intervals\ \cite{bryson1975applied}. We will
show that if $s_{n}\left(  t\right)  =a>0$ or $s_{n}\left(  t\right)  =b<L,$
then $\lambda_{s_{n}}\left(  t\right)  =0$ for some $n\in\left\{
1,\ldots,N\right\}  $ may in fact exist for some finite arc; otherwise
$\lambda_{s_{n}}\left(  t\right)  =0$ can arise only when $u_{n}\left(
t\right)  =0$.

The implication of (\ref{eq:multiDynCoI}) with $\lambda_{i}\left(  T\right)
=0$ is that $\lambda_{i}\left(  t\right)  =T-t$ for all $t\in\lbrack0,T]$ and
all $i=1,\ldots,M$ and that $\lambda_{i}\left(  t\right)  $ is monotonically
decreasing starting with $\lambda_{i}\left(  0\right)  =T$. However, this is
only true if the entire optimal trajectory is an interior arc, i.e., all
$R_{i}(t)\geq0$ constraints for all $i=1,\ldots,M$ remain inactive. On the
other hand, looking at (\ref{eq:multiDynCoS}), observe that when the two end
points, $0$ and $L$, are not within the range of an agent, we have $\left\vert
F_{n}^{-}(t)\right\vert =\left\vert F_{n}^{+}(t)\right\vert $, since the
number of indices $i$ satisfying $s_{n}\left(  t\right)  -r_{n}\leq\alpha
_{i}\leq s_{n}\left(  t\right)  $ is the same as that satisfying $s_{n}\left(
t\right)  <\alpha_{i}\leq s_{n}\left(  t\right)  +r_{n}$. Consequently, for
the one-agent case $N=1$, (\ref{eq:multiDynCoS}) becomes%
\begin{equation}
\dot{\lambda}_{s_{1}}\left(  t\right)  =-\frac{B}{r_{1}}\sum_{i\in F_{1}%
^{-}(t)}\lambda_{i}(t)+\frac{B}{r_{1}}\sum_{i\in F_{1}^{+}(t)}\lambda_{i}(t)
\label{eq:dynCoS}%
\end{equation}
and $\dot{\lambda}_{s_{1}}\left(  t\right)  =0$ since the two terms in
(\ref{eq:dynCoS}) will cancel out, i.e., $\lambda_{s_{1}}\left(  t\right)  $
remains constant as long as this condition is satisfied and, in addition, none
of the state constraints $R_{i}(t)\geq0$, $i=1,\ldots,M$, is active. Thus, for
the one agent case, as long as the optimal trajectory is an interior arc and
$\lambda_{s_{1}}\left(  t\right)  <0$, the agent moves at maximal speed
$u_{1}^{\star}\left(  t\right)  =1$ in the positive direction towards the
point $s_{1}=b$. If $\lambda_{s_{1}}\left(  t\right)  $ switches sign before
any of the state constraints $R_{i}(t)\geq0$, $i=1,\ldots,M$, becomes active
or the agent reaches the end point $s_{1}=b$, then $u_{1}^{\star}\left(
t\right)  =-1$ and the agent reverses its direction or, possibly, comes to rest.

In what follows, we examine the effect of the state constraints which
significantly complicates the analysis, leading to a challenging
two-point-boundary-value problem. However, we will establish the fact that the
complete solution boils down to determining a set of switching locations over
$[a,b]$ and waiting times at these switching points, with the end points, $0$
and $L$, being always infeasible on an optimal trajectory. This is a much
simpler problem that we are subsequently able to solve.

We begin by recalling that the dynamics in (\ref{eq:multiDynR}) indicate a
discontinuity arising when the condition $R_{i}(t)=0$ is satisfied while
$\dot{R}_{i}(t)=A_{i}-BP_{i}\left(  \mathbf{s}(t)\right)  <0$ for some
$i=1,\ldots,M$. Thus, $R_{i}=0$ defines an interior boundary condition which
is not an explicit function of time. Following standard optimal control
analysis \cite{bryson1975applied}, if this condition is satisfied at time $t$
for some $j\in\{1,\ldots,M\}$,
\begin{equation}
H\left(  \mathbf{x}(t^{-}),\mathbf{\lambda}(t^{-}),\mathbf{u}(t^{-})\right)
=H\left(  \mathbf{x}(t^{+}),\mathbf{\lambda}(t^{+}),\mathbf{u}(t^{+})\right)
\label{eq:hamcontinu}%
\end{equation}
where we note that one can choose to set the Hamiltonian to be continuous at
the entry point of a boundary arc or at the exit point. Using
(\ref{eq:Hamiltonian}) and (\ref{eq:multiDynR}), \eqref{eq:hamcontinu}
implies:%
\begin{equation}
\sum_{n=1}^{N}\lambda_{s_{n}}^{\ast}\left(  t^{-}\right)  u_{n}^{\ast}\left(
t^{-}\right)  +\lambda_{j}^{\star}\left(  t^{-}\right)  [A_{j}\left(
t\right)  -BP_{j}(\mathbf{s}(t))]=\sum_{n=1}^{N}\lambda_{s_{n}}^{\ast}\left(
t^{+}\right)  u_{n}^{\ast}\left(  t^{+}\right)  \label{eq:HamJumpR}%
\end{equation}
In addition, $\lambda_{s_{n}}^{\star}\left(  t^{-}\right)  =\lambda_{s_{n}%
}^{\star}\left(  t^{+}\right)  $ for all $n=$ $1,\ldots,N$ and $\lambda
_{i}^{\star}\left(  t^{-}\right)  =\lambda_{i}^{\star}\left(  t^{+}\right)  $
for all $i\neq j$, but $\lambda_{j}^{\star}\left(  t\right)  $ may experience
a discontinuity so that:%
\begin{equation}
\lambda_{j}^{\star}\left(  t^{-}\right)  =\lambda_{j}^{\star}\left(
t^{+}\right)  -\pi_{j} \label{eq:CostateDiscont}%
\end{equation}
where $\pi_{j}\geq0$ is a multiplier associated with the constraint
$-R_{j}(t)\leq0$. Recalling (\ref{eq:multiUstar}), since $\lambda_{s_{n}%
}^{\star}\left(  t\right)  $ remains unaffected, so does the optimal control,
i.e., $u_{n}^{\star}(t^{-})=u_{n}^{\star}(t^{+})$. Moreover, since this is an
entry point of a boundary arc, it follows from (\ref{eq:multiDynR}) that
$A_{j}-BP_{j}\left(  \mathbf{s}(t)\right)  <0$. Therefore, (\ref{eq:HamJumpR})
and (\ref{eq:CostateDiscont}) imply that
\[
\lambda_{j}^{\star}\left(  t^{-}\right)  =0,\text{ \ }\lambda_{j}^{\star
}\left(  t^{+}\right)  =\pi_{j}\geq0.
\]
Thus, $\lambda_{i}\left(  t\right)  $ always decreases with constant rate $-1$
until $R_{i}\left(  t\right)  =0$ is active, at which point $\lambda
_{i}\left(  t\right)  $ jumps to a non-negative value $\pi_{i}$ and decreases
with rate $-1$ again. The value of $\pi_{i}$ is determined by how long it
takes for the agents to reduce $R_{i}\left(  t\right)  $ to $0$ once again.
Obviously,
\begin{equation}
\lambda_{i}\left(  t\right)  \geq0,\text{ }i=1,\ldots,M\text{, }t\in\left[
0,T\right]  \label{eq:nonNegCoI}%
\end{equation}
with equality holding only if $t=T,$ or $t=$ $t_{0}^{-}$ with $R_{i}\left(
t_{0}\right)  =0$, $R_{i}\left(  t^{\prime}\right)  >0$, where $t^{\prime}%
\in\lbrack t_{0}-\delta,t_{0})$, $\delta>0.$ The actual evaluation of the
costate vector over the interval $[0,T]$ requires solving
(\ref{eq:multiDynCoS}), which in turn involves the determination of all points
where the state variables $R_{i}(t)$ reach their minimum feasible values
$R_{i}(t)=0$, $i=1,\ldots,M$. This generally involves the solution of a
two-point-boundary-value problem. However, our analysis thus far has already
established the structure of the optimal control (\ref{eq:multiUstar}) which
we have seen to remain unaffected by the presence of boundary arcs when
$R_{i}(t)=0$ for one or more $i=1,\ldots,M$. We will next prove some
additional structural properties of an optimal trajectory, based on which we
show that it is fully characterized by a set of non-negative scalar
parameters. Determining the values of these parameters is a much simpler
problem that does not require the solution of a two-point-boundary-value problem.

Let us turn our attention to the constraints $s_{n}(t)\geq a$ and
$s_{n}(t)\leq b$ and consider first the case where $a=0$, $b=L$, i.e., the
agents can move over the entire $[0,L]$. We shall make use of the following
technical condition:

%\textbf{Assumption 1}: For any $n=1,\ldots,N$, $i=1,\ldots,M$, $t\in(0,T)$,
%$\epsilon>0$, if $s_{n}(t)=0$, $s_{n}(t-\epsilon)>0$, then there doesn't exist
%$\delta\leq\epsilon,$ such that $R_{i}(t-\delta)>0$ and $R_{i}(t=0),$ for any
%$i=1,\ldots,M$; if $s_{n}(t)=L$, $s_{n}(t-\epsilon)<L$, then there doesn't
%exist $\delta\leq\epsilon,$ such that $R_{i}(t-\delta)>0$ and $R_{i}(t=0),$
%for any $i=1,\ldots,M$.

\textbf{Assumption 1}: For any $n=1,\ldots,N$, $i=1,\ldots,M$, $t\in(0,T)$,
and any $\epsilon>0$, if $s_{n}(t)=0$, $s_{n}(t-\epsilon)>0$, then either
$R_{i}(\tau)>0$ for all $\tau\in[t-\epsilon,t]$ or $R_{i}(\tau)=0$ for all
$\tau\in[t-\epsilon,t]$; if $s_{n}(t)=L, s_{n}(t-\epsilon)<L$,then either
$R_{i}(\tau)>0$ for all $\tau\in[t-\epsilon,t]$ or $R_{i}(\tau)=0$ for all
$\tau\in[t-\epsilon,t]$.

This condition excludes the case where an agent reaches an endpoint of the
mission space at the exact same time that any one of the uncertainty functions
reaches its minimal value of zero. Then, the following proposition asserts
that neither of the constraints $s_{n}(t)\geq0$ and $s_{n}(t)\leq L$ can
become active on an optimal trajectory.

\begin{proposition}
\label{lem:switchingpoints} Under Assumption 1, if $a=0$, $b=L$, then on an
optimal trajectory: $s_{n}^{\star}\left(  t\right)  \neq0$ and $s_{n}^{\star
}\left(  t\right)  \neq L$ for all $t\in(0,T)$, $n\in\left\{  1,\ldots
,N\right\}  .$
\end{proposition}

\textbf{Proof}. Suppose at $t=t_{0}<T$ an agent reaches the left endpoint,
i.e., $s_{n}^{\ast}\left(  t_{0}\right)  =0$, $s_{n}^{\ast}\left(  t_{0}%
^{-}\right)  >0$. We will then establish a contradiction. Thus, assuming
$s_{n}^{\ast}\left(  t_{0}\right)  =0$, we first show that $\lambda_{s_{n}%
}^{\ast}\left(  t_{0}^{-}\right)  =0$ by a contradiction argument. Assume that
$\lambda_{s_{n}}^{\ast}\left(  t_{0}^{-}\right)  \neq0$, in which case, since
the agent is moving toward $s_{n}=0$, we have $u_{n}^{\ast}\left(  t_{0}%
^{-}\right)  =-1$ and $\lambda_{s_{n}}^{\ast}\left(  t_{0}^{-}\right)  >0$
from (\ref{eq:multiUstar}). Then, $\lambda_{s_{n}}^{\ast}\left(  t\right)  $
may experience a discontinuity so that%
\begin{equation}
\lambda_{s_{n}}^{\ast}\left(  t_{0}^{-}\right)  =\lambda_{s_{n}}^{\ast}\left(
t_{0}^{+}\right)  -\pi_{n} \label{eq:costateDis}%
\end{equation}
where $\pi_{n}\geqslant0$ is a scalar constant. It follows that $\lambda
_{s_{n}}^{\ast}\left(  t_{0}^{+}\right)  =\lambda_{s_{n}}^{\ast}\left(
t_{0}^{-}\right)  +\pi_{n}>0$. Since the constraint $s_{n}\left(  t\right)
=0$ is not an explicit function of time, we have%
\begin{equation}
\lambda_{s_{n}}^{\ast}\left(  t_{0}^{-}\right)  u_{n}^{\ast}\left(  t_{0}%
^{-}\right)  =\lambda_{s_{n}}^{\ast}\left(  t_{0}^{+}\right)  u_{n}^{\ast
}\left(  t_{0}^{+}\right)  \label{eq:BoundaryCostate}%
\end{equation}
On the other hand, $u_{n}^{\ast}\left(  t_{0}^{+}\right)  \geqslant0$, since
agent $n$ must either come to rest or reverse its motion at $s_{n}=0$, hence
$\lambda_{s_{n}}^{\ast}\left(  t_{0}^{+}\right)  u_{n}^{\ast}\left(  t_{0}%
^{+}\right)  \geqslant0$. This violates (\ref{eq:BoundaryCostate}), since
$\lambda_{s_{n}}^{\ast}\left(  t_{0}^{-}\right)  u_{n}^{\ast}\left(  t_{0}%
^{-}\right)  <0.$ This contradiction implies that $\lambda_{s_{n}}^{\ast
}\left(  t_{0}^{-}\right)  =0$. Next, consider (\ref{eq:multiDynCoS}) and
observe that in (\ref{eq:multiFset}) we have $F_{n}^{-}\left(  t_{0}\right)
=\varnothing$, since $\alpha_{i}>s_{n}^{\ast}\left(  t_{0}\right)  =0$ for all
$i=1,\ldots,M$. Therefore, recalling (\ref{eq:nonNegCoI}), it follows from
(\ref{eq:multiDynCoS}) that%
\[
\dot{\lambda}_{s_{n}}\left(  t_{0}^{-}\right)  =\frac{B}{r_{n}}\sum
_{i\in\digamma_{n}^{+}\left(  t_{0}^{-}\right)  }\lambda_{i}\left(  t_{0}%
^{-}\right)
%TCIMACRO{\dprod \limits_{d\neq n}}%
%BeginExpansion
{\displaystyle\prod\limits_{d\neq n}}
%EndExpansion
\left[  1-p_{d,i}(s_{d}\left(  t_{0}^{-}\right)  )\right]  \geq0
\]
Under Assumption 1, there exists $\delta_{1}>0$ such that during the interval
$(t_{0}-\delta_{1},t_{0})$ no $R_{i}\left(  t\right)  \geq0$ becomes active,
hence no $\lambda_{i}(t)$ encounters a jump for $i=1,\ldots,M$. It follows
that $\lambda_{i}^{\ast}(t)>0$ for $i\in F_{n}^{+}(t)$ and $\dot{\lambda
}_{s_{n}}^{\ast}\left(  t\right)  $ is continuous with $\dot{\lambda}_{s_{n}%
}^{\ast}\left(  t\right)  >0$ for $t\in(t_{0}-\delta_{1},t_{0})$. Again, since
$s_{n}^{\ast}\left(  t_{0}\right)  =0,$ there exists some $\delta_{2}%
\leq\delta_{1}$ such that for $t\in\left(  t_{0}-\delta_{2},t_{0}\right)  $,
we have $u_{n}^{\ast}\left(  t\right)  <0$ and $\lambda_{s_{n}}^{\ast}\left(
t\right)  \geq0$. Thus, for $t\in\left(  t_{0}-\delta_{2},t_{0}\right)  $, we
have $\lambda_{s_{n}}^{\ast}\left(  t\right)  \geq0$ and $\dot{\lambda}%
_{s_{n}}^{\ast}\left(  t\right)  $ $>0$. This contradicts the fact we already
established that $\lambda_{s_{n}}^{\ast}\left(  t_{0}^{-}\right)  =0$ and we
conclude that $s_{n}^{\star}\left(  t\right)  \neq0$ for all $t\in\left[
0,T\right]  $, $n=1,\ldots,N$. Using a similar line of argument, we can also
show that $s_{n}^{\ast}\left(  t\right)  \neq L$. $\blacksquare$

\begin{proposition}
\label{lem:dwellSwitch} If $a>0$ and (or) $b<L$, then on an optimal trajectory
there exist finite length intervals $[t_{0},t_{1}]$ such that $s_{n}\left(
t\right)  =a$ and (or) $s_{n}\left(  t\right)  =b$, for some $n\in\left\{
1,\ldots,N\right\}  $, $t\in\lbrack t_{0},t_{1}]$, $0\leq t_{0}<$ $t_{1}\leq
T$.
\end{proposition}

\textbf{Proof}. Proceeding as in the proof of Proposition
\ref{lem:switchingpoints}, when $s_{n}^{\ast}\left(  t_{0}\right)  =a$ we can
establish (\ref{eq:BoundaryCostate}) and the fact that $\lambda_{s_{n}}^{\ast
}\left(  t_{0}^{-}\right)  =0$. On the other hand, $u_{n}^{\ast}\left(
t_{0}^{+}\right)  \geqslant0$, since the agent must either come to rest or
reverse its motion at $s_{n}\left(  t_{0}\right)  =a$. In other words, when
$s_{n}\left(  t_{0}\right)  =a$ on an optimal trajectory,
(\ref{eq:BoundaryCostate}) is satisfied either with the agent reversing its
direction immediately (in which case $t_{1}=t_{0}$ and $\lambda_{s_{n}}^{\ast
}\left(  t_{0}^{+}\right)  =0$) or staying on the boundary arc for a finite
time interval (in which case $t_{1}>t_{0}$ and $u_{n}^{\ast}\left(  t\right)
=0$ for $t\in\lbrack t_{0},t_{1}]$). The exact same argument can be applied to
$s_{n}\left(  t\right)  =b$. $\blacksquare$

The next result establishes the fact that on an optimal trajectory, every
agent either moves at full speed or is at rest.

\begin{proposition}
\label{lem:fullSpeedStop} On an optimal trajectory, either $u_{n}^{\ast
}\left(  t\right)  =\pm1$ if $\lambda_{s_{n}}^{\ast}\left(  t\right)  \neq0$,
or $u_{n}^{\ast}\left(  t\right)  =0$ if $\lambda_{s_{n}}^{\ast}\left(
t\right)  =0$ for $t\in\left[  0,T\right]  $, $n=1,\ldots,N$.
\end{proposition}

\textbf{Proof}. When $\lambda_{s_{n}}^{\ast}\left(  t\right)  \neq0$, we have
shown in (\ref{eq:multiUstar}) that $u_{n}^{\ast}\left(  t\right)  =\pm1$,
depending on the sign of $\lambda_{s_{n}}^{\ast}\left(  t\right)  $. Thus, it
remains to consider the case $\lambda_{s_{n}}^{\ast}\left(  t\right)  =0$ for
some $t\in\left[  t_{1},t_{2}\right]  $, where $0\leq t_{1}<t_{2}\leq T$.
Since the state is in a singular arc, $\lambda_{s_{n}}^{\ast}\left(  t\right)
$ does not provide information about $u_{n}^{\ast}\left(  t\right)  $. On the
other hand, the Hamiltonian in (\ref{eq:Hamiltonian}) is not a explicit
function of time, therefore, setting $H\left(  \mathbf{x}^{\star
},\mathbf{\lambda}^{\star},\mathbf{u}^{\star}\right)  \equiv H^{\ast}$, we
have$\frac{dH^{\ast}}{dt}=0$, which gives%
\begin{equation}
\frac{dH^{\ast}}{dt}=\sum_{i=1}^{M}\dot{R}_{i}^{\ast}(t)+\sum_{n=1}^{N}%
\dot{\lambda}_{s_{n}}^{\ast}\left(  t\right)  u_{n}^{\ast}\left(  t\right)
+\sum_{n=1}^{N}\lambda_{s_{n}}^{\ast}\left(  t\right)  \dot{u}_{n}^{\ast
}\left(  t\right)  +\sum_{i=1}^{M}\dot{\lambda}_{i}^{\ast}\left(  t\right)
\dot{R}_{i}^{\ast}(t)+\sum_{i=1}^{M}\lambda_{i}^{\ast}\left(  t\right)
\ddot{R}_{i}^{\ast}(t)=0 \label{eq:HtimeInvar1}%
\end{equation}
Define $S\left(  t\right)  =\left\{  n|\lambda_{s_{n}}\left(  t\right)
=0,n=1,\ldots,N\right\}  $ as the set of indices of agents that are in a
singular arc and $\bar{S}\left(  t\right)  =\left\{  n|\lambda_{s_{n}}\left(
t\right)  \neq0,n=1,\ldots,N\right\}  $ as the set of indices of all other
agents. Thus, $\lambda_{s_{n}}^{\ast}\left(  t\right)  =0$, $\dot{\lambda
}_{s_{n}}^{\ast}\left(  t\right)  =0$ for $t\in\left[  t_{1},t_{2}\right]
,n\in S\left(  t\right)  $. In addition, agents move with constant full speed,
either $1$ or $-1$, so that $\dot{u}_{n}^{\ast}\left(  t\right)  =0$,
$n\in\bar{S}\left(  t\right)  $. Then, (\ref{eq:HtimeInvar1}) becomes%
\begin{equation}
\frac{dH^{\ast}}{dt}=\sum_{i=1}^{M}[1+\dot{\lambda}_{i}^{\ast}\left(
t\right)  ]\dot{R}_{i}^{\ast}(t)+\sum_{n\in\bar{S}\left(  t\right)  }%
\dot{\lambda}_{s_{n}}^{\ast}\left(  t\right)  u_{n}^{\ast}\left(  t\right)
+\sum_{i=1}^{M}\lambda_{i}^{\ast}\left(  t\right)  \ddot{R}_{i}^{\ast}(t)=0
\label{eq:HtimeInvar2}%
\end{equation}
From (\ref{eq:multiDynCoI}), $\dot{\lambda}_{i}^{\ast}\left(  t\right)  =-1,$
$i=1,\ldots,M,$ so $1+\dot{\lambda}_{i}^{\ast}\left(  t\right)  =0$, leaving
only the last two terms above. Note that $\dot{\lambda}_{s_{n}}^{\ast}\left(
t\right)  =-\frac{\partial H^{\ast}}{\partial s_{n}^{\ast}\left(  t\right)  }$
and writing $\ddot{R}_{i}^{\ast}(t)=\frac{d\dot{R}_{i}^{\ast}(t)}{dt}$ we get:%
\[
-\sum_{n\in\bar{S}\left(  t\right)  }u_{n}^{\ast}\left(  t\right)
\frac{\partial H^{\ast}}{\partial s_{n}^{\ast}\left(  t\right)  }%
+\sum_{i=1,R_{i}\neq0}^{M}\lambda_{i}^{\ast}\left(  t\right)  \frac{d\dot
{R}_{i}^{\ast}(t)}{dt}=0
\]
Recall from (\ref{eq:multiDynR}) that when $R_{i}\left(  t\right)  \neq0$ we
have $\dot{R}_{i}\left(  t\right)  =A_{i}-B[1-%
%TCIMACRO{\dprod \limits_{n=1}^{N}}%
%BeginExpansion
{\displaystyle\prod\limits_{n=1}^{N}}
%EndExpansion
\left[  1-p_{i}(s_{n}\left(  t\right)  )\right]  ]$, so that%
\[
\frac{\partial H^{\ast}}{\partial s_{n}^{\ast}\left(  t\right)  }%
=-B\sum_{i=1,R_{i}\neq0}^{M}\lambda_{i}^{\ast}\left(  t\right)  \frac{\partial
p_{i}\left(  s_{n}^{\ast}\left(  t\right)  \right)  }{\partial s_{n}^{\ast
}\left(  t\right)  }%
%TCIMACRO{\dprod \limits_{d\neq n}^{N}}%
%BeginExpansion
{\displaystyle\prod\limits_{d\neq n}^{N}}
%EndExpansion
\left(  1-p_{i}\left(  s_{d}^{\ast}\left(  t\right)  \right)  \right)
\]%
\[
\frac{d\dot{R}_{i}^{\ast}(t)}{dt}=-B\sum_{n=1}^{N}u_{n}^{\ast}\left(
t\right)  \frac{\partial p_{i}\left(  s_{n}^{\ast}\left(  t\right)  \right)
}{\partial s_{n}^{\ast}\left(  t\right)  }%
%TCIMACRO{\dprod \limits_{d\neq n}^{N}}%
%BeginExpansion
{\displaystyle\prod\limits_{d\neq n}^{N}}
%EndExpansion
\left(  1-p_{i}\left(  s_{d}^{\ast}\left(  t\right)  \right)  \right)
\]
which results in
\begin{align}
&  B\sum_{i=1,R_{i}\neq0}^{M}\lambda_{i}^{\ast}\left(  t\right)  \left[
\sum_{n\in\bar{S}\left(  t\right)  }u_{n}^{\ast}\left(  t\right)
\frac{\partial p_{i}\left(  s_{n}^{\ast}\left(  t\right)  \right)  }{\partial
s_{n}^{\ast}\left(  t\right)  }%
%TCIMACRO{\dprod \limits_{d\neq n}^{N}}%
%BeginExpansion
{\displaystyle\prod\limits_{d\neq n}^{N}}
%EndExpansion
\left(  1-p_{i}\left(  s_{d}^{\ast}\left(  t\right)  \right)  \right)
-\sum_{n=1}^{N}u_{n}^{\ast}\left(  t\right)  \frac{\partial p_{i}\left(
s_{n}^{\ast}\left(  t\right)  \right)  }{\partial s_{n}^{\ast}\left(
t\right)  }%
%TCIMACRO{\dprod \limits_{d\neq n}^{N}}%
%BeginExpansion
{\displaystyle\prod\limits_{d\neq n}^{N}}
%EndExpansion
\left(  1-p_{i}\left(  s_{d}^{\ast}\left(  t\right)  \right)  \right)  \right]
\nonumber\\
&  =-B\sum_{i=1,R_{i}\neq0}^{M}\lambda_{i}^{\ast}\left(  t\right)  \sum_{n\in
S\left(  t\right)  }u_{n}^{\ast}\left(  t\right)  \frac{\partial p_{i}\left(
s_{n}^{\ast}\left(  t\right)  \right)  }{\partial s_{n}^{\ast}\left(
t\right)  }%
%TCIMACRO{\dprod \limits_{d\neq n}^{N}}%
%BeginExpansion
{\displaystyle\prod\limits_{d\neq n}^{N}}
%EndExpansion
\left(  1-p_{i}\left(  s_{d}^{\ast}\left(  t\right)  \right)  \right)  =0
\label{eq:HtimeInvar3}%
\end{align}
Note that $\frac{\partial p_{i}\left(  s_{1}^{\ast}\left(  t\right)  \right)
}{\partial s_{n}^{\ast}\left(  t\right)  }=\pm\frac{1}{r_{1}}$ or $0$,
depending on the relative position of $s_{1}^{\ast}\left(  t\right)  $ with
respect to $\alpha_{i}.$ Moreover, (\ref{eq:HtimeInvar3}) is invariant to $M$
or the precise way in which the mission space $[0,L]$ is partitioned, which
implies that
\[
\lambda_{i}^{\ast}\left(  t\right)  \sum_{n\in S\left(  t\right)  }u_{n}%
^{\ast}\left(  t\right)  \frac{\partial p_{i}\left(  s_{n}^{\ast}\left(
t\right)  \right)  }{\partial s_{n}^{\ast}\left(  t\right)  }%
%TCIMACRO{\dprod \limits_{d\neq n}^{N}}%
%BeginExpansion
{\displaystyle\prod\limits_{d\neq n}^{N}}
%EndExpansion
\left(  1-p_{i}\left(  s_{d}^{\ast}\left(  t\right)  \right)  \right)  =0
\]
for all $i=1,\ldots,M$, $t\in\left[  t_{1},t_{2}\right]  .$ Since
$\dot{\lambda}_{i}^{\ast}\left(  t\right)  =-1,$ $i=1,\ldots,M$, it is clear
that to satisfy this equality we must have $u_{n}^{\ast}\left(  t\right)  =0$
for all $t\in\left[  t_{1},t_{2}\right]  ,n\in S\left(  t\right)  $. In
conclusion, in a singular arc with $\lambda_{s_{n}}^{\ast}\left(  t\right)
=0$ for some $n\in\left\{  1,\ldots,N\right\}  ,$ the optimal control is
$u_{n}^{\ast}\left(  t\right)  =0$. $\blacksquare$

Next, we consider the case where the additional state constraint
(\ref{eq:multiNoCross}) is included. We can then prove that this constraint is
never active on an optimal trajectory, i.e., agents reverse their direction
before making contact with any other agent.

\begin{proposition}
\label{lem:noCross} If the constraint (\ref{eq:multiNoCross}) is included in
problem\textbf{ P1}, then on an optimal trajectory, $s_{n}^{\ast}\left(
t\right)  \neq s_{n+1}^{\ast}\left(  t\right)  $ for $t\in(0,T]$,
$n=1,\ldots,N-1$.
\end{proposition}

\textbf{Proof}. Suppose at $t=t_{0}<T$ we have $s_{n}^{\ast}\left(
t_{0}\right)  =s_{n+1}^{\ast}\left(  t_{0}\right)  $, for some $n=1,\ldots
,N-1.$ We will then establish a contradiction. First assuming that both agents
are moving (as opposed to one being at rest) toward each other, we have
$u_{n}^{\ast}\left(  t_{0}^{-}\right)  =1$ and $u_{n+1}^{\ast}\left(
t_{0}^{-}\right)  =-1.$ From (\ref{eq:multiUstar}) and Prop
\ref{lem:fullSpeedStop}, we know $\lambda_{s_{n}}^{\ast}\left(  t_{0}%
^{-}\right)  <0$ and $\lambda_{s_{n+1}}^{\ast}\left(  t_{0}^{-}\right)  >0.$
When the constraint $s_{n}\left(  t\right)  -s_{n+1}\left(  t\right)  \leq0$
is active, $\lambda_{s_{n}}^{\ast}\left(  t\right)  $ and $\lambda_{s_{n+1}%
}^{\ast}\left(  t_{0}^{-}\right)  $ may experience a discontinuity so that%
\begin{align}
\lambda_{s_{n}}^{\ast}\left(  t_{0}^{-}\right)   &  =\lambda_{s_{n}}^{\ast
}\left(  t_{0}^{+}\right)  +\pi\\
\lambda_{s_{n+1}}^{\ast}\left(  t_{0}^{-}\right)   &  =\lambda_{s_{n+1}}%
^{\ast}\left(  t_{0}^{+}\right)  -\pi
\end{align}
where $\pi\geqslant0$ is a scalar constant. It follows that $\lambda_{s_{n}%
}^{\ast}\left(  t_{0}^{+}\right)  =\lambda_{s_{n}}^{\ast}\left(  t_{0}%
^{-}\right)  -\pi<0$ and $\lambda_{s_{n+1}}^{\ast}\left(  t_{0}^{+}\right)
=\lambda_{s_{n+1}}^{\ast}\left(  t_{0}^{-}\right)  +\pi>0$. Since the
constraint $s_{n}\left(  t\right)  -s_{n+1}\left(  t\right)  \leq0$ is not an
explicit function of time, we have%
\begin{equation}
\lambda_{s_{n}}^{\ast}\left(  t_{0}^{-}\right)  u_{n}^{\ast}\left(  t_{0}%
^{-}\right)  +\lambda_{s_{n+1}}^{\ast}\left(  t_{0}^{-}\right)  u_{n+1}^{\ast
}\left(  t_{0}^{-}\right)  =\lambda_{s_{n}}^{\ast}\left(  t_{0}^{+}\right)
u_{n}^{\ast}\left(  t_{0}^{+}\right)  +\lambda_{s_{n+1}}^{\ast}\left(
t_{0}^{+}\right)  u_{n+1}^{\ast}\left(  t_{0}^{+}\right)
\label{eq:noCrossCos}%
\end{equation}
On the other hand, $u_{n}^{\ast}\left(  t_{0}^{+}\right)  \leqslant0$ and
$u_{n+1}^{\ast}\left(  t_{0}^{+}\right)  \geqslant0$, since agents $n$ and
$n+1$ must either come to rest or reverse their motion after making contact,
hence $\lambda_{s_{n}}^{\ast}\left(  t_{0}^{+}\right)  u_{n}^{\ast}\left(
t_{0}^{+}\right)  +\lambda_{s_{n+1}}^{\ast}\left(  t_{0}^{+}\right)
u_{n+1}^{\ast}\left(  t_{0}^{+}\right)  \geqslant0$. This violates
(\ref{eq:noCrossCos}), since $\lambda_{s_{n}}^{\ast}\left(  t_{0}^{-}\right)
u_{n}^{\ast}\left(  t_{0}^{-}\right)  +\lambda_{s_{n+1}}^{\ast}\left(
t_{0}^{-}\right)  u_{n+1}^{\ast}\left(  t_{0}^{-}\right)  <0.$ This
contradiction implies that $s_{n}\left(  t\right)  -s_{n+1}\left(  t\right)
=0$ cannot be active and we conclude that $s_{n}^{\ast}\left(  t\right)  \neq
s_{n+1}^{\ast}\left(  t\right)  $ for $t\in\left[  0,T\right]  $,
$n=1,\ldots,N-1$. Moreover, if one of the two agents is at rest when
$s_{n}^{\ast}\left(  t_{0}\right)  =s_{n+1}^{\ast}\left(  t_{0}\right)  $, the
same argument still holds since it is still true that $\lambda_{s_{n}}^{\ast
}\left(  t_{0}^{-}\right)  u_{n}^{\ast}\left(  t_{0}^{-}\right)
+\lambda_{s_{n+1}}^{\ast}\left(  t_{0}^{-}\right)  u_{n+1}^{\ast}\left(
t_{0}^{-}\right)  <0$. $\blacksquare$

Based on this analysis, the optimal control $u_{n}^{\ast}\left(  t\right)  $
depends entirely on the sign of $\lambda_{s_{n}}^{\ast}\left(  t\right)  $
and, in light of Propositions \ref{lem:switchingpoints}%
-\ref{lem:fullSpeedStop}, the solution of the problem reduces to determining:
$(i)$ \emph{switching points} in $[0,L]$ where an agent switches from
$u_{n}^{\ast}\left(  t\right)  =\pm1$ to either $\mp1$ or $0$; or from
$u_{n}^{\ast}\left(  t\right)  =0$ to either $\pm1$, and $(ii)$ if an agent
switches from $u_{n}^{\ast}\left(  t\right)  =\pm1$ to $0$, \emph{waiting
times} until the agent switches back to a speed $u_{n}^{\ast}\left(  t\right)
=\pm1$. In other words, the full solution is characterized by two parameter
vectors for each agent $n$: $\theta_{n}=[\theta_{n,1},\ldots,\theta
_{n,\Gamma_{n}}]^{\mathtt{T}}$ and $w_{n}=[w_{n,1}\ldots,w_{n,\Gamma_{n}%
}]^{\text{T}}$, where $\theta_{n,\xi}\in(0,L)$ denotes the $\xi$th location
where agent $n$ changes its speed from $\pm1$ to $0$ and $w_{n,\xi}\geq0$
denotes the time (which is possibly null) that agent $n$ dwells on
$\theta_{n,\xi}$. Note that $\Gamma_{n}$ is generally not known a priori and
depends on the time horizon $T$. In addition, we always assume that agent $n$
reverses its velocity direction after leaving the switching point
$\theta_{n,\xi}$ with respect to the one it had when reaching $\theta_{n,\xi}%
$. This seemingly excludes the possibility of an agent's control following a
sequence $1,0,1$ or $-1,0,-1$. However, these two motion behaviors can be
captured as two adjacent switching points approaching each other: when
$\left\vert \theta_{n,\xi}-\theta_{n,\xi+1}\right\vert \rightarrow0$, the
agent control follows the sequence $1,0,1$ or $-1,0,-1$, and the waiting time
associated with $u_{n}^{\ast}\left(  t\right)  =0$ is $w_{n,\xi}+w_{n,\xi+1}.$

For simplicity, we will assume that $s_{n}(0)=0$, so that it follows from
Proposition \ref{lem:switchingpoints} that $u_{n}^{\star}(0)=1$, $n=$
$1,\ldots,N$. Therefore, $\theta_{n,1}$ corresponds to the optimal control
switching from $1$ to $0$. Furthermore, $\theta_{n,\xi}$ with $\xi$ odd (even)
always corresponds to $u_{n}^{\star}(t)$ switching from $1$ to $0$ ($-1$ to
$0$.) Thus, we have the following constraints on the switching locations for
all $\xi=2,\ldots,\Gamma_{n}$:%
\begin{equation}
\left\{
\begin{array}
[c]{l}%
\theta_{n,\xi}\leq\theta_{n,\xi-1},\text{ if }\xi\text{ is even}\\
\theta_{n,\xi}\geq\theta_{n,\xi-1},\text{ if }\xi\text{ is odd}.
\end{array}
\right.  \label{eq:thetaconstraint}%
\end{equation}

%$0\leq\theta_{1}\leq...\leq \theta_{N}\leq T$ and the number of switches, $N$, depends on the time horizon $T$ considered.  Without the loss of generality, we assume that all  Thus, the optimal control
%problem reduces to a nonlinear optimization problem with controllable
%parameters $\theta_{1},\ldots,\theta_{N}$. We also observe that the
%sequence of time points $\{\theta_{i}\}$ can be readily transformed to
%a sequence of switching locations $\{z_{i}\}$, $i=1,\ldots,N$, which, by Lem. \ref{lem:switchingpoints}, satisfies
%$0<z_{i}<L$ for all $i=1,\ldots,N$. Given that the agent's optimal speed
%is always constant ($1$ or $-1$), it is easy to check that for all
%$n=1,2,\ldots$:
%\begin{eqnarray}
%\label{eq:timetolocation}
%z_{2n} &=&2\sum_{i=1}^{n}\theta_{2i-1}-2\sum_{i=1}^{n-1}\theta_{2i}-\theta_{2n}-s(0)\nonumber\\
%z_{2n+1} &=&2\sum_{i=1}^{n}\theta_{2i-1}-2\sum_{i=1}^{n}\theta_{2i}+\theta_{2n}-s(0).
%\end{eqnarray}
It is now clear that the behavior of each agent under the optimal control
policy is that of a \emph{hybrid system} whose dynamics undergo switches when
$u_{n}^{\star}\left(  t\right)  $ changes from $\pm1$ to $0$ and from $0$ to
$\mp1$ or when $R_{i}(t)$ reaches or leaves the boundary value $R_{i}=0$. As a
result, we are faced with a parametric optimization problem for a system with
hybrid dynamics. This is a setting where one can apply the generalized theory
of Infinitesimal Perturbation Analysis (IPA) in
\cite{cassandras2009perturbation},\cite{Wardietal09} to conveniently obtain
the gradient of the objective function $J$ in (\ref{eq:costfunction}) with
respect to the vectors $\theta$ and $w$, and therefore, determine (generally,
locally) optimal vectors $\theta^{\star}$ and $w^{\ast}$ through a
gradient-based optimization approach. Note that this is done on line, i.e.,
the gradient is evaluated by observing a trajectory with given $\theta$ and
$w$ over $[0,T]$ based on which $\theta$ and $w$ are adjusted until
convergence is attained using standard gradient-based algorithms.

\textbf{Remark 1}. If the agent dynamics in (\ref{eq:multiDynOfS}) are
replaced by a model such as $\dot{s}_{n}(t)=g_{n}(s_{n})+b_{n}u_{n}(t)$,
observe that (\ref{eq:multiUstar}) still holds. The difference lies in
(\ref{eq:multiDynCoS}) which would involve a dependence on $\frac{dg_{n}%
(s_{n})}{ds_{n}}$ and further complicate the associated
two-point-boundary-value problem. However, since the optimal solution is also
defined by a parameter vectors $\theta_{n}=[\theta_{n,1},\ldots,\theta
_{n,\Gamma_{n}}]^{\mathtt{T}}$ and $w_{n}=[w_{n,1}\ldots,w_{n,\Gamma_{n}%
}]^{\text{T}}$ for each agent $n$, we can still apply the IPA approach
presented in the next section.

\subsection{Infinitesimal Perturbation Analysis (IPA)}

\label{sec:IPA}

Our analysis thus far has shown that, on an optimal trajectory, the agent
moves at full speed, dwells on a switching point (possibly for zero time) and
never reaches either boundary point, i.e., $0<s_{n}^{\star}(t)<L$. Thus, the
$n$th agent's movement can be parameterized through $\theta_{n}=[\theta
_{n,1},\ldots,\theta_{n,\Gamma_{n}}]^{\mathtt{T}}$ and $w_{n}=[w_{n,1}%
\ldots,w_{n,\Gamma_{n}}]^{\text{T}}$ where $\theta_{n,\xi}$ is the $\xi$th
control switching point and $w_{n,\xi}$ is the waiting time for this agent at
the $\xi$th switching point. Therefore, the solution of problem \textbf{P1}
reduces to the determination of optimal parameter vectors $\theta_{n}^{\star}$
and $w_{n}^{\ast}$, $n=1,\ldots,N$. As we pointed out, the agent's optimal
behavior defines a hybrid system, and the switching locations translate to
switching times between particular modes of this system. This is similar to
switching-time optimization problems, e.g., \cite{egerstedt2006transition}
,\cite{shaikh2007hybrid},\cite{xu2004optimal}, except that we can only control
a subset of mode switching times. We make use of IPA in part to exploit
robustness properties that the resulting gradients possess
\cite{yaoc2011perturbation}; specifically, we will show that they do not
depend on the uncertainty model parameters $A_{i}$, $i=1,\ldots,M$, and may
therefore be used without any detailed knowledge of how uncertainty affects
the mission space.

\subsubsection{One agent solution with $a=0$ and $b=L$}

\label{sec:oneagent}

To maintain some notational simplicity, we begin with a single agent who can
move on the entire mission space $[0,L]$ and will then provide the natural
extension to multiple agents and a mission space limited to $[a,b]\subset
\lbrack0,L]$. We present the associated hybrid automaton model for this
single-agent system operating on an optimal trajectory. Our goal is to
determine $\nabla J(\theta,w)$, the gradient of the objective function $J$ in
(\ref{eq:costfunction}) with respect to $\theta$ and $w$, which can then be
used in a gradient-based algorithm to obtain optimal parameter vectors
$\theta_{n}^{\star}$ and $w_{n}^{\ast}$, $n=1,\ldots,N$. We will apply IPA,
which provides a formal way to obtain state and event time derivatives with
respect to parameters of hybrid systems, from which we can subsequently
obtaining $\nabla J(\theta,w)$.
%Thus, we begin with a description of this system.
%However, since we are only interested at switching locations (and thus translates directly to times) where the control changes its sign, and not time instant when the system undergoes
%However, we choose to use the IPA approach to better exploit the structure of the problem. {\bf Dennis: We probably need a better explanation of why are we using IPA instead of these switching-time optimization approaches.  I think one reason is that since the hybrid system is very complex, it is hard to generate the right sequence of modes}

\textbf{Hybrid automaton model}. We use a standard definition of a hybrid
automaton (e.g., see \cite{cassandras2007stochastic}) as the formalism to
model the system described above. Thus, let $q\in Q$ (a countable set) denote
the discrete state (or mode) and $x\in X\subseteq\mathbb{R}^{n}$ denote the
continuous state. Let $\upsilon\in\Upsilon$ (a countable set) denote a
discrete control input and $u\in U\subseteq\mathbb{R}^{m}$ a continuous
control input. Similarly, let $\delta\in\Delta$ (a countable set) denote a
discrete disturbance input and $d\in D\subseteq\mathbb{R}^{p}$ a continuous
disturbance input. The state evolution is determined by means of $(i)$ a
vector field $f:Q\times X\times U\times D\rightarrow X$, $(ii)$ an invariant
(or domain) set $Inv:$ $Q\times\Upsilon\times\Delta\rightarrow2^{X}$, $(iii)$
a guard set $Guard:$ $Q\times Q\times\Upsilon\times\Delta\rightarrow2^{X}$,
and $(iv)$ a reset function $r:$ $Q\times Q\times X\times\Upsilon\times
\Delta\rightarrow X$. The system remains at a discrete state $q$ as long as
the continuous (time-driven) state $x$ does not leave the set $Inv(q,\upsilon
,\delta)$. If $x$ reaches a set $Guard(q,q^{\prime},\upsilon,\delta)$ for some
$q^{\prime}\in Q$, a discrete transition can take place. If this transition
does take place, the state instantaneously resets to $(q^{\prime},x^{\prime})$
where $x^{\prime}$ is determined by the reset map $r(q,q^{\prime}%
,x,\upsilon,\delta)$. Changes in $\upsilon$ and $\delta$ are discrete events
that either \emph{enable} a transition from $q$ to $q^{\prime}$ by making sure
$x\in Guard(q,q^{\prime},\upsilon,\delta)$ or \emph{force} a transition out of
$q$ by making sure $x\notin Inv(q,\upsilon,\delta)$. We will classify all
events that cause discrete state transitions in a manner that suits the
purposes of IPA. Since our problem is set in a deterministic framework,
$\delta$ and $d$ will not be used.

\begin{figure}[ptb]
\centering
\includegraphics[
height=4in,
width=6in
]{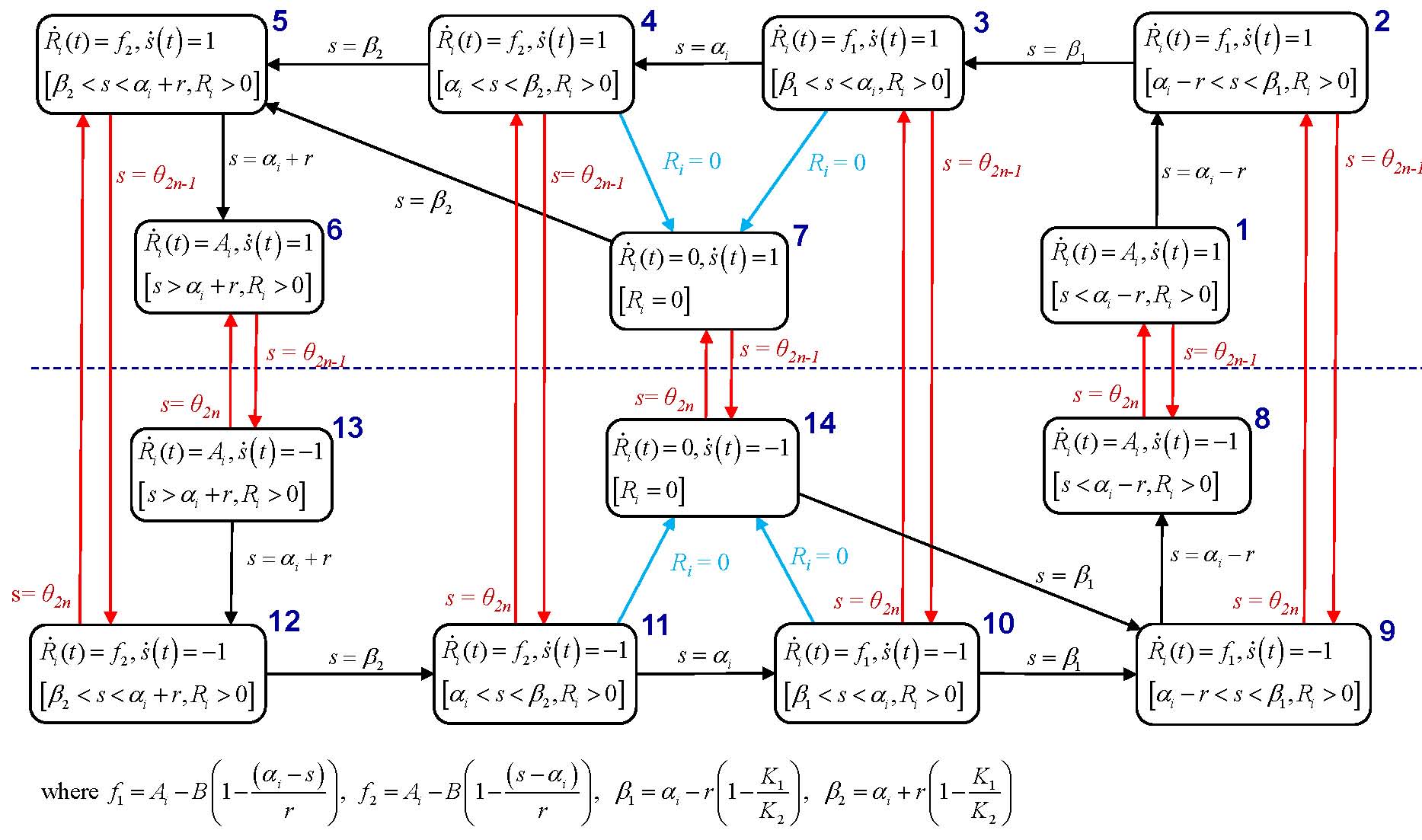}\caption{Hybrid automaton for each $\alpha_{i}$. Red arrows
represent events when the control switches between $1$ and $-1$. Blue arrows
represent events when $R_{i}$ becomes $0$. Black arrows represent all other
events.}%
\label{fig:stateTran}%
\end{figure}

We show in Fig. \ref{fig:stateTran} a partial hybrid automaton model of the
single-agent system where $a=0$ and $b=L$. Since there is only one agent, we
set $s\left(  t\right)  =s_{1}\left(  t\right)  $, $u\left(  t\right)
=u_{1}\left(  t\right)  $ and $\theta=\theta_{1}$ for simplicity. Due to the
size of the overall model, Fig. \ref{fig:stateTran} is limited to the behavior
of the agent with respect to a single $\alpha_{i},i\in\{1,\ldots,M\}$ and
ignores modes where the agent dwells on the switching points (these, however,
are included in our extended analysis in Section \ref{sec:multiagent}.) The
model consists of 14 discrete states (modes) and is symmetric in the sense
that states $1-7$ correspond to the agent operating with $u(t)=1$, and states
$8-14$ correspond to the agent operating with $u(t)=-1$. States where
$u\left(  t\right)  =0$ are omitted since we do not include the waiting time
parameter $w=w_{1}$ here$.$ The events that cause state transitions can be
placed in three categories: $(i)$ The value of $R_{i}(t)$ becomes 0 and
triggers a switch in the dynamics of (\ref{eq:multiDynR}). This can only
happen when $R_{i}(t)>0$ and $\dot{R}_{i}(t)=A_{i}-Bp_{i}(s(t))<0$ (e.g., in
states $3$ and $4$), causing a transition to state $7$ in which the invariant
condition is $R_{i}(t)=0$. $(ii)$ The agent reaches a switching location,
indicated by the guard condition $s(t)=\theta_{\xi}$ for any $\xi
=1,\ldots,\Gamma$. In these cases, a transition results from a state $z$ to
$z+7$ if $z=1,\ldots,6$ and to $z-7$ otherwise. $(iii)$ The agent position
reaches one of several critical values that affect the dynamics of $R_{i}(t)$
while $R_{i}(t)>0$. Specifically, when $s(t)=\alpha_{i}-r$, the value of
$p_{i}(s(t))$ becomes strictly positive and $\dot{R}_{i}(t)=A_{i}%
-Bp_{i}(s(t))>0$, as in the transition $1\rightarrow2$. Subsequently, when
$s(t)=\alpha_{i}-r(1-A_{i}/B)$, as in the transition $2\rightarrow3$, the
value of $p_{i}(s(t))$ becomes sufficiently large to cause $\dot{R}%
_{i}(t)=A_{i}-Bp_{i}(s(t))<0$ so that a transition due to $R_{i}(t)=0$ becomes
feasible at this state. Similar transitions occur when $s(t)=\alpha_{i}$,
$s(t)=\alpha_{i}+r(1-A_{i}/B)$, and $s(t)=\alpha_{i}+r$. The latter results in
state $6$ where $\dot{R}_{i}(t)=A_{i}>0$ and the only feasible event is
$s(t)=\theta_{\xi}$, $\xi$ odd, when a switch must occur and a transition to
state $13$ takes place (similarly for state $8$).

\textbf{IPA review}. Before proceeding, we provide a brief review of the IPA
framework for general stochastic hybrid systems as presented in
\cite{cassandras2009perturbation}. The purpose of IPA is to study the behavior
of a hybrid system state as a function of a parameter vector $\theta\in\Theta$
for a given compact, convex set $\Theta\subset\mathbb{R}^{l}$. Let $\{\tau
_{k}(\theta)\}$, $k=1,\ldots,K$, denote the occurrence times of all events in
the state trajectory. For convenience, we set $\tau_{0}=0$ and $\tau_{K+1}=T$.
Over an interval $[\tau_{k}(\theta),\tau_{k+1}(\theta))$, the system is at
some mode during which the time-driven state satisfies $\dot{x}\ =\ f_{k}%
(x,\theta,t)$. An event at $\tau_{k}$ is classified as $(i)$ \emph{Exogenous}
if it causes a discrete state transition independent of $\theta$ and satisfies
$\frac{d\tau_{k}}{d\theta}=0$; $(ii)$ \emph{Endogenous}, if there exists a
continuously differentiable function $g_{k}:\mathbb{R}^{n}\times
\Theta\rightarrow\mathbb{R}$ such that $\tau_{k}\ =\ \min\{t>\tau
_{k-1}\ :\ g_{k}\left(  x\left(  \theta,t\right)  ,\theta\right)  =0\}$; and
$(iii)$ \emph{Induced} if it is triggered by the occurrence of another event
at time $\tau_{m}\leq\tau_{k}$. IPA specifies how changes in $\theta$
influence the state $x(\theta,t)$ and the event times $\tau_{k}(\theta)$ and,
ultimately, how they influence interesting performance metrics which are
generally expressed in terms of these variables.

Given $\theta=[\theta_{1},\ldots,\theta_{\Gamma}]^{\mathtt{T}}$, we use the
Jacobian matrix notation: $x^{\prime}(t)\equiv\frac{\partial x(\theta
,t)}{\partial\theta}$, $\tau_{k}^{\prime}\equiv\frac{\partial\tau_{k}(\theta
)}{\partial\theta}$, $k=1,\ldots,K$, for all state and event time derivatives.
It is shown in \cite{cassandras2009perturbation} that $x^{\prime}(t)$
satisfies:%
\begin{equation}
\frac{d}{dt}x^{\prime}\left(  t\right)  =\frac{\partial f_{k}\left(  t\right)
}{\partial x}x^{\prime}\left(  t\right)  +\frac{\partial f_{k}\left(
t\right)  }{\partial\theta} \label{eq:xpDyn}%
\end{equation}
for $t\in\lbrack\tau_{k},\tau_{k+1})$ with boundary condition:
\begin{equation}
x^{\prime}(\tau_{k}^{+})\ =\ x^{\prime}(\tau_{k}^{-})+\left[  f_{k-1}(\tau
_{k}^{-})-f_{k}(\tau_{k}^{+})\right]  \tau_{k}^{\prime} \label{eq:xpBoundary}%
\end{equation}
for $k=0,\ldots,K$. In addition, in (\ref{eq:xpBoundary}), the gradient vector
for each $\tau_{k}$ is $\tau_{k}^{\prime}=0$ if the event at $\tau_{k}$ is
exogenous and
\begin{equation}
\tau_{k}^{\prime}=-\left[  \frac{\partial g_{k}}{\partial x}f_{k}(\tau_{k}%
^{-})\right]  ^{-1}\left(  \frac{\partial g_{k}}{\partial\theta}%
+\frac{\partial g_{k}}{\partial x}x^{\prime}(\tau_{k}^{-})\right)
\label{eq:taukp}%
\end{equation}
if the event at $\tau_{k}$ is endogenous (i.e., $\ g_{k}\left(  x\left(
\theta,\tau_{k}\right)  ,\theta\right)  =0$) and defined as long as
$\frac{\partial g_{k}}{\partial x}f_{k}(\tau_{k}^{-})\neq0$.

\textbf{IPA equations}. To clarify the presentation, we first note that
$i=1,\ldots,M$ is used to index the points where uncertainty is measured;
$\xi=1,\ldots,\Gamma$ indexes the components of the parameter vector; and
$k=1,\ldots,K$ indexes event times. In order to apply the three fundamental
IPA equations (\ref{eq:xpDyn})-(\ref{eq:taukp}) to our system, we use the
state vector $x\left(  t\right)  =[s\left(  t\right)  ,R_{1}(t),\ldots
,R_{M}(t)]^{\mathtt{T}}$ and parameter vector $\theta=[\theta_{1}%
,\ldots,\theta_{\Gamma}]^{\mathtt{T}}$. We then identify all events that can
occur in Fig. \ref{fig:stateTran} and consider intervals $[\tau_{k}%
(\theta),\tau_{k+1}(\theta))$ over which the system is in one of the 14 states
shown for each $i=1,\ldots,M$. Applying (\ref{eq:xpDyn}) to $s(t)$ with
$f_{k}\left(  t\right)  =1$ or $-1$ due to (\ref{eq:multiDynOfS}) and
(\ref{eq:multiUstar}), the solution yields the gradient vector $\nabla
s(t)=[\frac{\partial s}{\partial\theta_{1}}(t),\ldots,\frac{\partial
s}{\partial\theta_{M}}(t)]^{\mathtt{T}}$, where
\begin{equation}
\frac{\partial s}{\partial\theta_{\xi}}(t)=\frac{\partial s}{\partial
\theta_{\xi}}(\tau_{k}^{+}),\text{ for }t\in\lbrack\tau_{k},\tau_{k+1})
\label{eq:sprimet}%
\end{equation}
for all $k=1,\ldots,K$, i.e., for all states $z(t)\in\{1,\ldots,14\}$.
Similarly, let $\nabla R_{i}(t)=[\frac{\partial R_{i}}{\partial\theta_{1}%
}(t),\ldots,\frac{\partial R_{i}}{\partial\theta_{M}}(t)]^{\mathtt{T}}$ for
$i=1,\ldots,M$. We note from (\ref{eq:multiDynR}) that $f_{k}\left(  t\right)
=0$ for states $z(t)\in Z_{1}\equiv\{7,14\}$; $f_{k}\left(  t\right)  =A_{i}$
for states $z(t)\in Z_{2}\equiv\{1,6,8,13\}$; and $f_{k}\left(  t\right)
=A_{i}-Bp_{i}(s(t))$ for all other states which we further classify into
$Z_{3}\equiv\{2,3,11,12\}$ and $Z_{4}\equiv\{4,5,9,10\}$. Thus, solving
(\ref{eq:xpDyn}) and using (\ref{eq:sprimet}) gives:
\[
\nabla R_{i}\left(  t\right)  =\nabla R_{i}(\tau_{k}^{+})-\left\{
\begin{array}
[c]{ll}%
0 & \text{if }z\left(  t\right)  \in Z_{1}\cup Z_{2}\\
B\left(  \frac{\partial p_{i}(s)}{\partial s}\right)  \nabla s\left(  \tau
_{k}^{+}\right)  \cdot(t-\tau_{k}) & \text{otherwise}%
\end{array}
\right.
\]
where $\frac{\partial p_{i}(s)}{\partial s}=\pm\frac{1}{r}$ as evaluated from
(\ref{eq:multiLinearModel}) depending on the sign of $\alpha_{i}-s(t)$ at each
associated automaton state.

We now turn our attention to the determination of $\nabla s\left(  \tau
_{k}^{+}\right)  $ and $\nabla R_{i}(\tau_{k}^{+})$ which are needed to
evaluate $\nabla R_{i}\left(  t\right)  $ above. To do so, we use
(\ref{eq:xpBoundary}), which involves the event time gradient vectors
$\nabla\tau_{k}=[\frac{\partial\tau_{k}}{\partial\theta_{1}},\ldots
,\frac{\partial\tau_{k}}{\partial\theta_{\Gamma}}]^{\mathtt{T}}$ for
$k=1,\ldots,K$ (the value of $K$ depends on $T$.)
%To keep notation manageable in what follows, we define%
%\[
%s_{j}^{\prime}=\frac{\partial s\left(  t\right)  }{\partial\theta_{j}},\text{
%}R_{i,j}^{\prime}\left(  t\right)  =\frac{\partial \nabla R_{i}\left(
%t\right)  }{\partial\theta_{j}},\text{ }\tau_{k,j}^{\prime}=\frac{\partial
%\tau_{k}}{\partial\theta_{j}}%
%\]
Looking at Fig. \ref{fig:stateTran}, there are three readily distinguishable
cases regarding the events that cause discrete state transitions:

\emph{Case 1}: An event at time $\tau_{k}$ which is neither $R_{i}=0$ nor
$s=\theta_{\xi}$, for any $\xi=1,\ldots,\Gamma$. In this case, it is easy to
see that the dynamics of both $s(t)$ and $R_{i}(t)$ are continuous, so that
$f_{k-1}(\tau_{k}^{-})=f_{k}(\tau_{k}^{+})$ in (\ref{eq:xpBoundary}) applied
to $s\left(  t\right)  $ and $R_{i}(t),$ $i=1,\ldots,M$ gives:
\begin{equation}
\left\{
\begin{array}
[c]{l}%
\nabla s\left(  \tau_{k}^{+}\right)  =\nabla s\left(  \tau_{k}^{-}\right) \\
\nabla R_{i}(\tau_{k}^{+})=\nabla R_{i}(\tau_{k}^{-}),\text{ \ }i=1,\ldots,M
\end{array}
\right.
\end{equation}

\emph{Case 2}: An event $R_{i}=0$ at time $\tau_{k}$. This corresponds to
transitions $3\rightarrow7$, $4\rightarrow7$, $10\rightarrow14$ and
$11\rightarrow14$ in Fig. \ref{fig:stateTran} where the dynamics of $s(t)$ are
still continuous, but the dynamics of $R_{i}(t)$ switch from $f_{k-1}(\tau
_{k}^{-})=A_{i}-Bp_{i}(s(\tau_{k}^{-}))$ to $f_{k}(\tau_{k}^{+})=0$. Thus,
$\nabla s\left(  \tau_{k}^{-}\right)  =\nabla s\left(  \tau_{k}^{+}\right)  $,
but we need to evaluate $\tau_{k}^{\prime}$ to determine $\nabla R_{i}%
(\tau_{k}^{+})$. Observing that this event is endogenous, (\ref{eq:taukp})
applies with $g_{k}=R_{i}=0$ and we get%
\[
\frac{\partial\tau_{k}}{\partial\theta_{\xi}}=-\frac{\frac{\partial R_{i}%
}{\partial\theta_{\xi}}\left(  \tau_{k}^{-}\right)  }{A_{i}-Bp_{i}(s(\tau
_{k}^{-}))},\text{ \ \ }\xi=1,\ldots,\Gamma,\text{ }k=1,\ldots,K
\]
It follows from (\ref{eq:xpBoundary}) that%
\[
\frac{\partial R_{i}}{\partial\theta_{\xi}}\left(  \tau_{k}^{+}\right)
=\frac{\partial R_{i}}{\partial\theta_{\xi}}\left(  \tau_{k}^{-}\right)
-\frac{[A_{i}-Bp_{i}(s(\tau_{k}^{-}))]\frac{\partial R_{i}}{\partial
\theta_{\xi}}\left(  \tau_{k}^{-}\right)  }{A_{i}-Bp_{i}(s(\tau_{k}^{-}))}=0
\]
Thus, whenever an event occurs at $\tau_{k}$ such that $R_{i}(\tau_{k})$
becomes zero, $\frac{\partial R_{i}}{\partial\theta_{\xi}}\left(  \tau_{k}%
^{+}\right)  $ is always reset to $0$ regardless of $\frac{\partial R_{i}%
}{\partial\theta_{\xi}}\left(  \tau_{k}^{-}\right)  $.

\emph{Case 3}: An event at time $\tau_{k}$ due to a control sign change at
$s=\theta_{\xi}$, $\xi=1,\ldots,\Gamma$. This corresponds to any transition
between the upper and lower part of the hybrid automaton in Fig.
\ref{fig:stateTran}. In this case, the dynamics of $R_{i}(t)$ are continuous
and we have $\frac{\partial R_{i}}{\partial\theta_{\xi}}\left(  \tau_{k}%
^{+}\right)  =\frac{\partial R_{i}}{\partial\theta_{\xi}}\left(  \tau_{k}%
^{-}\right)  $ for all $i,\xi,k$. On the other hand, we have $\dot{s}(\tau
_{k}^{+})=u(\tau_{k}^{+})=-u(\tau_{k}^{-})=\pm1$. Observing that any such
event is endogenous, (\ref{eq:taukp}) applies with $g_{k}=s-\theta_{\xi}=0$
for some $\xi=1,\ldots,\Gamma$ and we get%
\begin{equation}
\frac{\partial\tau_{k}}{\partial\theta_{\xi}}=\frac{1-\frac{\partial
s}{\partial\theta_{\xi}}\left(  \tau_{k}^{-}\right)  }{u(\tau_{k}^{-})}
\label{eq:taukpN}%
\end{equation}
Combining (\ref{eq:taukpN}) with (\ref{eq:xpBoundary}) and recalling that
$u(\tau_{k}^{+})=-u(\tau_{k}^{-})$, we have%
\[
\frac{\partial s}{\partial\theta_{\xi}}(\tau_{k}^{+})=\ \frac{\partial
s}{\partial\theta_{\xi}}(\tau_{k}^{-})+[u\left(  \tau_{k}^{-}\right)
-u(\tau_{k}^{+})]\frac{1-\frac{\partial s}{\partial\theta_{\xi}}\left(
\tau_{k}^{-}\right)  }{u(\tau_{k}^{-})}=2
\]
where $\frac{\partial s}{\partial\theta_{\xi}}\left(  \tau_{k}^{-}\right)  =0$
because $\frac{\partial s}{\partial\theta_{\xi}}\left(  0\right)
=0=\frac{\partial s}{\partial\theta_{\xi}}\left(  t\right)  $ for all
$t\in\lbrack0,\tau_{k})$, since the position of the agent cannot be affected
by $\theta_{\xi}$ prior to this event.

In this case, we also need to consider the effect of perturbations to
$\theta_{j}$ for $j<\xi$, i.e., prior to the current event time $\tau_{k}$
(clearly, for $j>\xi$, $\frac{\partial s}{\partial\theta_{j}}(\tau_{k}^{+})=0$
since the current position of the agent cannot be affected by future events.)
Observe that since $g_{k}=s-\theta_{\xi}=0$, we have $\frac{\partial g_{k}%
}{\partial\theta_{j}}=0$ for $j\neq\xi$ and (\ref{eq:taukp}) gives%
\[
\frac{\partial\tau_{k}}{\partial\theta_{j}}=-\frac{\frac{\partial s}%
{\partial\theta_{j}}\left(  \tau_{k}^{-}\right)  }{u(\tau_{k}^{-})}%
\]
so that using this in (\ref{eq:xpBoundary}) we get:%
\[
\frac{\partial s}{\partial\theta_{j}}(\tau_{k}^{+})=\frac{\partial s}%
{\partial\theta_{j}}(\tau_{k}^{-})-\frac{\left[  u\left(  \tau_{k}^{-}\right)
-u(\tau_{k}^{+})\right]  \frac{\partial s}{\partial\theta_{j}}\left(  \tau
_{k}^{-}\right)  }{u\left(  \tau_{k}^{-}\right)  }=-\frac{\partial s}%
{\partial\theta_{j}}\left(  \tau_{k}^{-}\right)
\]
Combining the above results, the components of $\nabla s(\tau_{k}^{+})$ where
$\tau_{k}$ is the event time when $s(\tau_{k})=\theta_{\xi}$ for some $\xi$,
are given by
\begin{equation}
\frac{\partial s}{\partial\theta_{j}}(\tau_{k}^{+})\ =\left\{
\begin{array}
[c]{ll}%
-\frac{\partial s}{\partial\theta_{j}}\left(  \tau_{k}^{-}\right)  & \text{if
}j=1,\ldots,\xi-1\\
2 & \text{if }j=\xi\\
0 & \text{if }j>\xi
\end{array}
\right.  \label{eq:spJump}%
\end{equation}

It follows from (\ref{eq:sprimet}) and the analysis of all three cases above
that $\frac{\partial s}{\partial\theta_{\xi}}\left(  t\right)  $ for all $\xi$
is constant throughout an optimal trajectory except at transitions caused by
control switching locations (\emph{Case 3}). In particular, for the $k$th
event corresponding to $s(\tau_{k})=\theta_{\xi}$, $t\in\lbrack\tau_{k},T]$,
if $u\left(  t\right)  =1$, then $\frac{\partial s}{\partial\theta_{\xi}%
}\left(  t\right)  =-2$ if $\xi$ is odd, and $\frac{\partial s}{\partial
\theta_{\xi}}\left(  t\right)  =2$ if $\xi$ is even; similarly, if $u\left(
t\right)  =-1$, then $\frac{\partial s}{\partial\theta_{\xi}}\left(  t\right)
=2$ if $\xi$ is odd and $\frac{\partial s}{\partial\theta_{\xi}}\left(
t\right)  =-2$ if $\xi$ is even. In summary, we can write:
\begin{equation}
\frac{\partial s}{\partial\theta_{\xi}}\left(  t\right)  =\left\{
\begin{array}
[c]{cc}%
\left(  -1\right)  ^{\xi}\cdot2u\left(  t\right)  & t\geq\tau_{k}\\
0 & t<\tau_{k}%
\end{array}
\right.  \text{, \ }\xi=1,\ldots,\Gamma\label{eq:spTheta}%
\end{equation}
Finally, we can combine (\ref{eq:spTheta}) with our results for $\frac
{\partial R_{i}}{\partial\theta_{\xi}}\left(  t\right)  $ in all three cases
above. Letting $s(\tau_{l})=\theta_{\xi}$, we obtain the following expression
for $\frac{\partial R_{i}}{\partial\theta_{\xi}}\left(  t\right)  $ for all
$k\geq l,$ $t\in\lbrack\tau_{k},\tau_{k+1})$:
\begin{equation}
\frac{\partial R_{i}}{\partial\theta_{\xi}}\left(  t\right)  =\frac{\partial
R_{i}}{\partial\theta_{\xi}}\left(  \tau_{k}^{+}\right)  +\left\{
\begin{array}
[c]{cl}%
0 & \text{if }z(t)\in Z_{1}\cup Z_{2}\\
\left(  -1\right)  ^{\xi+1}\frac{2B}{r}u\left(  \tau_{k}^{+}\right)
\cdot(t-\tau_{k}) & \text{if }z(t)\in Z_{3}\\
-\left(  -1\right)  ^{\xi+1}\frac{2B}{r}u\left(  \tau_{k}^{+}\right)
\cdot(t-\tau_{k}) & \text{if }z(t)\in Z_{4}%
\end{array}
\right.  \label{eq:RpDyn1}%
\end{equation}
with boundary condition
\begin{equation}
\frac{\partial R_{i}}{\partial\theta_{\xi}}(\tau_{k}^{+})=\left\{
\begin{array}
[c]{cl}%
0 & \text{if }z\left(  \tau_{k}^{+}\right)  \in Z_{1}\\
\frac{\partial R_{i}}{\partial\theta_{\xi}}(\tau_{k}^{-}) & \text{otherwise}%
\end{array}
\right.  \label{eq:RpJump}%
\end{equation}

\textbf{Objective Function Gradient Evaluation.} Based on our analysis, the
objective function (\ref{eq:costfunction}) in problem \textbf{P1} can now be
written as $J(\theta)$, a function of $\theta$ instead of $u\left(  t\right)
$ and we can rewrite it as%
\[
J(\theta)=\frac{1}{T}\sum_{i=1}^{M}\sum_{k=0}^{K}%
%TCIMACRO{\dint _{\tau_{k}(\theta)}^{\tau_{k+1}(\theta)}}%
%BeginExpansion
{\displaystyle\int_{\tau_{k}(\theta)}^{\tau_{k+1}(\theta)}}
%EndExpansion
R_{i}\left(  t,\theta\right)  dt
\]
where we have explicitly indicated the dependence on $\theta$. We then
obtain:
\[
\nabla J(\theta)=\frac{1}{T}%
%TCIMACRO{\dsum _{i=1}^{M}}%
%BeginExpansion
{\displaystyle\sum_{i=1}^{M}}
%EndExpansion%
%TCIMACRO{\dsum _{k=0}^{K}}%
%BeginExpansion
{\displaystyle\sum_{k=0}^{K}}
%EndExpansion
\left(
%TCIMACRO{\dint _{\tau_{k}}^{\tau_{k+1}}}%
%BeginExpansion
{\displaystyle\int_{\tau_{k}}^{\tau_{k+1}}}
%EndExpansion
\nabla R_{i}\left(  t\right)  dt+R_{i}\left(  \tau_{k+1}\right)  \nabla
\tau_{k+1}-R_{i}\left(  \tau_{k}\right)  \nabla\tau_{k}\right)
\]
Observing the cancelation of all terms of the form $R_{i}\left(  \tau
_{k}\right)  \nabla\tau_{k}$ for all $k$ (with $\tau_{0}=0$, $\tau_{K+1}=T$
fixed), we finally get
\begin{equation}
\nabla J(\theta)=\frac{1}{T}%
%TCIMACRO{\dsum _{i=1}^{M}}%
%BeginExpansion
{\displaystyle\sum_{i=1}^{M}}
%EndExpansion%
%TCIMACRO{\dsum _{k=0}^{K}}%
%BeginExpansion
{\displaystyle\sum_{k=0}^{K}}
%EndExpansion%
%TCIMACRO{\dint _{\tau_{k}(\theta)}^{\tau_{k+1}(\theta)}}%
%BeginExpansion
{\displaystyle\int_{\tau_{k}(\theta)}^{\tau_{k+1}(\theta)}}
%EndExpansion
\nabla R_{i}\left(  t\right)  dt. \label{eq:dJdTheta}%
\end{equation}
The evaluation of $\nabla J(\theta)$ therefore depends entirely on $\nabla
R_{i}\left(  t\right)  $, which is obtained from (\ref{eq:RpDyn1}%
)-(\ref{eq:RpJump}) and the event times $\tau_{k}$, $k=1,\ldots,K$, given
initial conditions $s\left(  0\right)  =0$, $R_{i}\left(  0\right)  $ for
$i=1,\ldots,M$ and $\nabla R_{i}(0)=0$. Since $\nabla R_{i}\left(  t\right)  $
itself depends only on the event times $\tau_{k}$, $k=1,\ldots,K$, the
gradient $\nabla J(\theta)$ is obtained by observing the switching times in a
trajectory over $[0,T]$ characterized by the vector $\theta$.

\subsubsection{Multi agent solution where $a\geq0$ and $b\leq L$}

\label{sec:multiagent}

Next, we extend the results obtained in the previous section to the general
multi-agent problem where we also allow $a\geq0$ and $b\leq L$. Recall that we
require $0\leq a\leq r_{n}$ and $L-r_{m}\leq b\leq L$, for at least some
$n,m=1,\ldots,N$ since, otherwise, controlling agent movement cannot affect
$R_{i}(t)$ for all $\alpha_{i}$ located outside the sensing range of agents.
We now include both parameter vectors $\theta_{n}=[\theta_{n,1},\ldots
,\theta_{n,\Gamma_{n}}]^{\mathtt{T}}$ and $w_{n}=[w_{n,1},\ldots
w_{n,\Gamma_{n}}]^{\text{T}}$ for each agent $n$ and, for notational
simplicity, concatenate them to construct $\theta=\left[  \theta_{1}%
,\ldots,\theta_{N}\right]  ^{\mathtt{T}}$ and $w=\left[  w_{1},\ldots
,w_{N}\right]  ^{\mathtt{T}}$. The solution of problem \textbf{P1} reduces to
the determination of optimal parameter vectors $\theta^{\star}$ and $w^{\ast}$
and we will use IPA to evaluate $\nabla J(\theta,w)=[\frac{dJ\left(
\theta,w\right)  }{d\theta}\frac{dJ\left(  \theta,w\right)  }{dw}%
]^{\mathtt{T}}$. Similar to (\ref{eq:dJdTheta}), it is clear that this depends
on $\nabla R_{i}(t)=\left[  \frac{\partial R_{i}\left(  t\right)  }%
{\partial\theta}\frac{\partial R_{i}\left(  t\right)  }{\partial w}\right]
^{\mathtt{T}}$ and the event times $\tau_{k}$, $k=1,\ldots,K,$ observed on a
trajectory over $[0,T]$ with given $\theta$ and $w$.

\textbf{IPA equations}. We begin by recalling the dynamics of $R_{i}\left(
t\right)  $ in (\ref{eq:multiDynR}) which depend on the relative positions of
all agents with respect to $\alpha_{i}$ and change at time instants $\tau_{k}$
such that either $R_{i}(\tau_{k})=0$ with $R_{i}(\tau_{k}^{-})>0$ or
$A_{i}>BP_{i}\left(  \mathbf{s}(\tau_{k})\right)  $ with $R_{i}(\tau_{k}%
^{-})=0$. Moreover, using (\ref{eq:multiDynOfS}) and our analysis in Section
\ref{sec:Hamiltonian}, the dynamics of $s_{n}\left(  t\right)  $,
$n=1,\ldots,N$, in an optimal trajectory can be expressed as follows. Define
$\Theta_{n,\xi}=(\theta_{n,\xi-1},\theta_{n,\xi})$ if $\xi$ is odd and
$\Theta_{n,\xi}=(\theta_{n,\xi},\theta_{n,\xi-1})$ if $\xi$ is even to be the
$\xi$th interval between successive switching points for any $n=1,\ldots,N$,
where $\theta_{n,0}=s_{n}(0)$. Then, for $\xi=1,2,\ldots$,
\begin{equation}
\dot{s}_{n}\left(  t\right)  =\left\{
\begin{array}
[c]{ll}%
1 & s_{n}(t)\in\Theta_{n,\xi},\text{ }\xi\text{ odd}\\
-1 & s_{n}(t)\in\Theta_{n,\xi},\text{ }\xi\text{ even}\\
0 & \text{otherwise}%
\end{array}
\right.  \label{eq:optdyns}%
\end{equation}
where transitions for $s_{n}\left(  t\right)  $ from $\pm1$ to $\mp1$ are
incorporated by treating them as cases where $w_{n,\xi}=0$, i.e., no dwelling
at a switching point $\theta_{n,\xi}$ (in which case $\dot{s}_{n}\left(
t\right)  =0$.) We can now concentrate on all events causing switches either
in the dynamics of any $R_{i}\left(  t\right)  $, $i=1,\ldots,M$, or the
dynamics of any $s_{n}(t)$, $n=1,\ldots,N$. From (\ref{eq:xpBoundary}), any
other event at some time $\tau_{k}$ in this hybrid system cannot modify the
values of $\nabla R_{i}(t)=\left[  \frac{\partial R_{i}\left(  t\right)
}{\partial\theta}\frac{\partial R_{i}\left(  t\right)  }{\partial w}\right]
^{\mathtt{T}}$ or $\nabla s_{n}(t)=\left[  \frac{\partial s_{n}\left(
t\right)  }{\partial\theta_{n}}\frac{\partial s_{n}\left(  t\right)
}{\partial w_{n}}\right]  ^{\mathtt{T}}$ at $t=\tau_{k}$.

First, applying (\ref{eq:xpDyn}) to $s_{n}(t)$ with $f_{k}\left(  t\right)
=1$, $-1$ or $0$ due to (\ref{eq:optdyns}), the solution yields%
\begin{equation}
\nabla s_{n}(t)=\nabla s_{n}(\tau_{k}^{+}),\text{ for }t\in\lbrack\tau
_{k},\tau_{k+1}) \label{eq:wSpt}%
\end{equation}
for all $k=1,\ldots,K$, $n=$ $1,\ldots,N.$ Similarly, applying (\ref{eq:xpDyn}%
) to $R_{i}\left(  t\right)  $ and using (\ref{eq:multiDynR}) gives:
\begin{equation}
\frac{\partial R_{i}}{\partial\theta_{n,\xi}}\left(  t\right)  =\frac{\partial
R_{i}}{\partial\theta_{n,\xi}}\left(  \tau_{k}^{+}\right)  -\left\{
\begin{array}
[c]{ll}%
0 & \text{if }R_{i}(t)=0,\text{ }A_{i}<BP_{i}\left(  \mathbf{s}(t)\right) \\
B%
%TCIMACRO{\dprod \limits_{d\neq n}}%
%BeginExpansion
{\displaystyle\prod\limits_{d\neq n}}
%EndExpansion
\left(  1-p_{i}\left(  s_{d}\left(  t\right)  \right)  \right)  \left(
\frac{\partial p_{i}(s_{n})}{\partial s_{n}}\right)  \frac{\partial
s_{n}\left(  \tau_{k}^{+}\right)  }{\partial\theta_{n,\xi}}\cdot(t-\tau_{k}) &
\text{otherwise}%
\end{array}
\right.  \label{eq:sSpt}%
\end{equation}
and%
\begin{equation}
\frac{\partial R_{i}}{\partial w_{n,\xi}}\left(  t\right)  =\frac{\partial
R_{i}}{\partial w_{n,\xi}}\left(  \tau_{k}^{+}\right)  -\left\{
\begin{array}
[c]{ll}%
0 & \text{if }R_{i}(t)=0,\text{ }A_{i}<BP_{i}\left(  \mathbf{s}(t)\right) \\
B%
%TCIMACRO{\dprod \limits_{d\neq n}}%
%BeginExpansion
{\displaystyle\prod\limits_{d\neq n}}
%EndExpansion
\left(  1-p_{i}\left(  s_{d}\left(  t\right)  \right)  \right)  \left(
\frac{\partial p_{i}(s_{n})}{\partial s_{n}}\right)  \frac{\partial
s_{n}\left(  \tau_{k}^{+}\right)  }{\partial w_{n,\xi}}\cdot(t-\tau_{k}) &
\text{otherwise}%
\end{array}
\right.  \label{eq:wSpt1}%
\end{equation}
Thus, it remains to determine the components of $\nabla s_{n}\left(  \tau
_{k}^{+}\right)  $ and $\nabla R_{i}(\tau_{k}^{+})$ in (\ref{eq:wSpt}%
)-(\ref{eq:wSpt1}) using (\ref{eq:xpBoundary}). This involves the event time
gradient vectors $\nabla\tau_{k}=\left[  \frac{\partial\tau_{k}}%
{\partial\theta}\frac{\partial\tau_{k}}{\partial w}\right]  ^{\mathtt{T}}$ for
$k=1,\ldots,K$, which will be determined through (\ref{eq:taukp}). There are
three possible cases regarding the events that cause switches in the dynamics
of $R_{i}\left(  t\right)  $ or $s_{n}(t)$ as mentioned above:

\emph{Case 1}: An event at time $\tau_{k}$ such that $\dot{R}_{i}\left(
t\right)  $ switches from $\dot{R}_{i}\left(  t\right)  =0$ to $\dot{R}%
_{i}\left(  t\right)  =A_{i}-BP_{i}\left(  \mathbf{s}(t)\right)  $. In this
case, it is easy to see that the dynamics of both $s_{n}(t)$ and $R_{i}(t)$
are continuous, so that $f_{k-1}(\tau_{k}^{-})=f_{k}(\tau_{k}^{+})$ in
(\ref{eq:xpBoundary}) applied to $s_{n}\left(  t\right)  $ and $R_{i}(t),$
$i=1,\ldots,M$, $n=1,\ldots,N$, and we get%
\begin{align}
\nabla s_{n}\left(  \tau_{k}^{+}\right)   &  =\nabla s_{n}\left(  \tau_{k}%
^{-}\right)  ,\text{ }n=1,\ldots,N\label{eq:wsn}\\
\nabla R_{i}(\tau_{k}^{+})  &  =\nabla R_{i}(\tau_{k}^{-}),\text{ }%
i=1,\ldots,M \label{eq:wRi}%
\end{align}

\emph{Case 2}: An event at time $\tau_{k}$ such that $\dot{R}_{i}\left(
t\right)  $ switches from $\dot{R}_{i}\left(  t\right)  =A_{i}-BP_{i}\left(
\mathbf{s}(t)\right)  $ to $\dot{R}_{i}\left(  t\right)  =0$, i.e.,
$R_{i}(\tau_{k})$ becomes zero. In this case, we need to first evaluate
$\nabla\tau_{k}$ from (\ref{eq:taukp}) in order to determine $\nabla
R_{i}(\tau_{k}^{+})$ through (\ref{eq:xpBoundary}). Observing that this event
is endogenous, (\ref{eq:taukp}) applies with $g_{k}=R_{i}=0$ and we get%
\begin{equation}
\nabla\tau_{k}=-\frac{\nabla R_{i}(\tau_{k}^{-})}{A_{i}\left(  \tau_{k}%
^{-}\right)  -BP_{i}\left(  \mathbf{s}(\tau_{k}^{-})\right)  }
\label{eq:wtaukpR0}%
\end{equation}
It follows from (\ref{eq:xpBoundary}) that%
\begin{equation}
\nabla R_{i}(\tau_{k}^{+})=\nabla R_{i}(\tau_{k}^{-})-\frac{[A_{i}\left(
\tau_{k}^{-}\right)  -BP_{i}\left(  \mathbf{s}(t)\right)  ]\nabla R_{i}\left(
\tau_{k}^{-}\right)  }{A_{i}\left(  \tau_{k}^{-}\right)  -BP_{i}\left(
\tau_{k}^{-}\right)  }=0 \label{eq:wRpR0}%
\end{equation}
Thus, $\nabla R_{i}(\tau_{k}^{+})$ is always reset to $0$ regardless of
$\nabla R_{i}(\tau_{k}^{-})$. In addition, (\ref{eq:wsn}) holds, since the the
dynamics of $s_{n}(t)$ are continuous at time $\tau_{k}$.

\emph{Case 3}: An event at time $\tau_{k}$ such that the dynamics of
$s_{n}\left(  t\right)  $ switch from $\pm1$ to $0$, or from $0$ to $\pm1$.
Clearly, (\ref{eq:wRi}) holds since the the dynamics of $R_{i}(t)$ are
continuous at this time. However, determining $\nabla s_{n}\left(  \tau
_{k}^{+}\right)  $ is more elaborate and requires us to consider its
components separately, first $\frac{\partial s_{n}\left(  \tau_{k}^{+}\right)
}{\partial\theta_{n}}$ and then $\frac{\partial s_{n}\left(  \tau_{k}%
^{+}\right)  }{\partial w_{n}}$.

\emph{Case 3.1: }Evaluation of\emph{ }$\frac{\partial s_{n}\left(  \tau
_{k}^{+}\right)  }{\partial\theta_{n}}$.

\emph{Case 3.1.1: }An event at time $\tau_{k}$ such that the dynamics of
$s_{n}(t)$ in (\ref{eq:optdyns}) switch from $\pm1$ to $0$. This is an
endogenous event and (\ref{eq:taukp}) applies with $g_{k}=s_{n}-\theta_{n,\xi
}=0$ for some $\xi=1,\ldots,\Gamma_{n}$ and we have:%
\begin{equation}
\frac{\partial\tau_{k}}{\partial\theta_{n,\xi}}=\frac{1-\frac{\partial s_{n}%
}{\partial\theta_{n,\xi}}\left(  \tau_{k}^{-}\right)  }{u_{n}(\tau_{k}^{-})}
\label{eq:wtaukpThetaComeToRest}%
\end{equation}
and (\ref{eq:xpBoundary}) yields%
\begin{equation}
\frac{\partial s_{n}}{\partial\theta_{n,\xi}}(\tau_{k}^{+})=\ \frac{\partial
s_{n}}{\partial\theta_{n,\xi}}(\tau_{k}^{-})+[u_{n}\left(  \tau_{k}%
^{-}\right)  -0]\frac{1-\frac{\partial s_{n}}{\partial\theta_{n,\xi}}\left(
\tau_{k}^{-}\right)  }{u_{n}(\tau_{k}^{-})}=1 \label{eq:wSpthetaComeToRest}%
\end{equation}
As in Case 3 of Section \ref{sec:oneagent}, we also need to consider the
effect of perturbations to $\theta_{j}$ for $j<\xi$, i.e., prior to the
current event time $\tau_{k}$ (clearly, for $j>\xi$, $\frac{\partial s_{n}%
}{\partial\theta_{j}}(\tau_{k}^{+})=0$ since the current position of the agent
cannot be affected by future events.) Observe that $\frac{\partial g_{k}%
}{\partial\theta_{j}}=0$, therefore, (\ref{eq:taukp}) becomes%
\begin{equation}
\frac{\partial\tau_{k}}{\partial\theta_{n,j}}=-\frac{\frac{\partial s_{n}%
}{\partial\theta_{n,j}}\left(  \tau_{k}^{-}\right)  }{u_{n}(\tau_{k}^{-})}
\label{eq:wtaukpThetaComeToRestAft}%
\end{equation}
and using this in (\ref{eq:xpBoundary}) gives:%
\begin{equation}
\frac{\partial s_{n}}{\partial\theta_{n,j}}(\tau_{k}^{+})=\frac{\partial
s_{n}}{\partial\theta_{n,j}}(\tau_{k}^{-})-\frac{\left[  u_{n}\left(  \tau
_{k}^{-}\right)  -0\right]  \frac{\partial s_{n}}{\partial\theta_{n,j}}\left(
\tau_{k}^{-}\right)  }{u_{n}\left(  \tau_{k}^{-}\right)  }=0
\label{eq:wSpThetaComeToRestAft}%
\end{equation}
Thus, combining the above results, when $s_{q}(\tau_{k})=\theta_{q,\xi}$ for
some $\xi$ and the agent switches from $\pm1$ to $0$, we have
\begin{equation}
\frac{\partial s_{n}}{\partial\theta_{n,j}}(\tau_{k}^{+})\ =\left\{
\begin{array}
[c]{ll}%
0, & \text{if }j\neq\xi\\
1, & \text{if }j=\xi
\end{array}
\right.  \label{eq:wSpJump}%
\end{equation}
\emph{Case 3.1.2: }An event at time $\tau_{k}$ such that the dynamics of
$s_{n}(t)$ in (\ref{eq:optdyns}) switch from $0$ to $\pm1$. This is an induced
event since it is\ triggered by the occurrence of some other endogenous event
when the agent switches from $\pm1$ to $0$ (see \emph{Case 3.1.1} above.)
Suppose the agent starts from an initial position $s_{n}\left(  0\right)  =a$
with $u_{n}\left(  0\right)  =1$ and $\tau_{k}$ is the time the agent switches
from the $0$ to $\pm1$ at the switching point $\theta_{n,\xi}.$ If
$\theta_{n,\xi}$ is such that $u_{n}\left(  \tau_{k}^{+}\right)  =1$, then
$\xi$ is even and $\tau_{k}$ can be calculated as follows:%
\begin{align}
\tau_{k}  &  =(\theta_{n,1}-a)+w_{n,1}+\left(  \theta_{n,1}-\theta
_{n,2}\right)  +w_{n,2}+\ldots+\left(  \theta_{n,\xi-1}-\theta_{n,\xi}\right)
+w_{n,\xi}\label{eq:wtaukZeroToOneL}\\
&  =2\left(
%TCIMACRO{\dsum \limits_{v=1,\text{ }v\text{ odd}}^{\xi-1}}%
%BeginExpansion
{\displaystyle\sum\limits_{v=1,\text{ }v\text{ odd}}^{\xi-1}}
%EndExpansion
\theta_{n,v}-\sum_{v=2,\text{ }v\text{ even}}^{\xi-2}\theta_{n,v}\right)  +%
%TCIMACRO{\dsum \limits_{v=1}^{\xi}}%
%BeginExpansion
{\displaystyle\sum\limits_{v=1}^{\xi}}
%EndExpansion
w_{n,v}-\theta_{n,\xi}\nonumber
\end{align}
Similarly, if $\theta_{n,\xi}$ is the switching point such that $u_{n}\left(
\tau_{k}^{+}\right)  =-1$, then $\xi$ is odd and we get:%

\begin{equation}
\tau_{k}=2\left(
%TCIMACRO{\dsum \limits_{v=1,\text{ }v\text{ odd}}^{\xi-2}}%
%BeginExpansion
{\displaystyle\sum\limits_{v=1,\text{ }v\text{ odd}}^{\xi-2}}
%EndExpansion
\theta_{n,v}-\sum_{v=2,\text{ }v\text{ even}}^{\xi-1}\theta_{n,v}\right)  +%
%TCIMACRO{\dsum \limits_{v=1}^{\xi}}%
%BeginExpansion
{\displaystyle\sum\limits_{v=1}^{\xi}}
%EndExpansion
w_{n,v}+\theta_{n,\xi} \label{eq:wtaukZeroToOneR2}%
\end{equation}
We can then directly obtain $\frac{\partial\tau_{k}}{\partial\theta_{n,\xi}}$
as%
\begin{equation}
\frac{\partial\tau_{k}}{\partial\theta_{n,\xi}}=-sgn(u\left(  \tau_{k}%
^{+}\right)  ) \label{eq:wtaukpZeroToOne}%
\end{equation}
Using (\ref{eq:wtaukpZeroToOne}) in (\ref{eq:xpBoundary}) gives:%
\begin{equation}
\frac{\partial s_{n}}{\partial\theta_{n,\xi}}(\tau_{k}^{+})=\frac{\partial
s_{n}}{\partial\theta_{n,\xi}}(\tau_{k}^{-})+\left[  0-u\left(  \tau_{k}%
^{+}\right)  \right]  \cdot\lbrack-sgn(u\left(  \tau_{k}^{+}\right)
)]=\frac{\partial s_{n}}{\partial\theta_{n,\xi}}(\tau_{k}^{-})+1
\label{eq:wSpZeroToOne}%
\end{equation}
Once again, we need to consider the effect of perturbations to $\theta_{j}$
for $j<\xi$, i.e., prior to the current event time $\tau_{k}$ (clearly, for
$j>\xi$, $\frac{\partial s_{n}}{\partial\theta_{j}}(\tau_{k}^{+})=0$.) In this
case, from (\ref{eq:wtaukZeroToOneL})-(\ref{eq:wtaukZeroToOneR2}), we have%
\begin{equation}
\left\{
\begin{array}
[c]{cl}%
\frac{\partial\tau_{k}}{\partial\theta_{n,j}}=2, & \text{if }j\text{ odd}\\
\frac{\partial\tau_{k}}{\partial\theta_{n,j}}=-2, & \text{if }j\text{ even}%
\end{array}
\right.  \label{eq:wtaukpZeroToOne2}%
\end{equation}
and it follows from (\ref{eq:xpBoundary}) that for $j<\xi$:%
\begin{equation}
\frac{\partial s_{n}}{\partial\theta_{n,j}}(\tau_{k}^{+})=\left\{
\begin{array}
[c]{cl}%
\frac{\partial s_{n}}{\partial\theta_{n,j}}(\tau_{k}^{-})+2, & \text{if }%
u_{n}\left(  \tau_{k}^{+}\right)  =1,\text{ }j\text{ even, or }u_{n}\left(
\tau_{k}^{+}\right)  =-1,\text{ }j\text{ odd}\\
\frac{\partial s_{n}}{\partial\theta_{n,j}}(\tau_{k}^{-})-2, & \text{if }%
u_{n}\left(  \tau_{k}^{+}\right)  =1,\text{ }j\text{ odd, or }u_{n}\left(
\tau_{k}^{+}\right)  =-1,\text{ }j\text{ even}%
\end{array}
\right.  \label{eq:wSpZeroToOne2}%
\end{equation}

\emph{Case 3.2:} Evaluation of\emph{ } $\frac{\partial s_{n}\left(  \tau
_{k}^{+}\right)  }{\partial w_{n}}.$

\emph{Case 3.2.1: }An event at time $\tau_{k}$ such that the dynamics of
$s_{n}(t)$ in (\ref{eq:optdyns}) switch from $\pm1$ to $0$. This is an
endogenous event and (\ref{eq:taukp}) applies with $g_{k}=s_{n}-\theta_{n,\xi
}=0$ for some $\xi=1,\ldots,\Gamma_{n}$. Then, for any $j\leq\xi$, we have:
\begin{equation}
\frac{\partial\tau_{k}}{\partial w_{n,j}}=\frac{-\frac{\partial s_{n}%
}{\partial w_{n,j}}\left(  \tau_{k}^{-}\right)  }{u_{n}(\tau_{k}^{-})}
\label{eq:wtaukpW}%
\end{equation}
Combining (\ref{eq:wtaukpW}) with (\ref{eq:xpBoundary}) and since
$u_{n}\left(  \tau_{k}^{-}\right)  =$ $\pm1$, we have%
\begin{equation}
\frac{\partial s_{n}}{\partial w_{n,j}}(\tau_{k}^{+})=\frac{\partial s_{n}%
}{\partial w_{n,j}}(\tau_{k}^{-})+[u_{n}\left(  \tau_{k}^{-}\right)
-0]\frac{-\frac{\partial s_{n}}{\partial w_{n,j}}\left(  \tau_{k}^{-}\right)
}{u_{n}(\tau_{k}^{-})}=0 \label{eq:wSpJumpW2}%
\end{equation}

\emph{Case 3.2.2: }An event at time $\tau_{k}$ such that the dynamics of
$s_{n}(t)$ in (\ref{eq:optdyns}) switch from $0$ to $\pm1$. As in \emph{Case
3.1.2}, $\tau_{k}$ is given by (\ref{eq:wtaukZeroToOneL}) or
(\ref{eq:wtaukZeroToOneR2}), depending on the sign of $u_{q}\left(  \tau
_{k}^{+}\right)  $. Thus, we have $\frac{\partial\tau_{k}}{\partial w_{n,j}%
}=1$, for $j\leq\xi$. Using this result in (\ref{eq:xpBoundary}) and observing
that $\frac{\partial s_{n}}{\partial w_{n,j}}(\tau_{k}^{-})=0$ from
(\ref{eq:wSpJumpW2}), we have%
\begin{equation}
\frac{\partial s_{n}}{\partial w_{n,j}}(\tau_{k}^{+})=\ \frac{\partial s_{n}%
}{\partial w_{n,j}}(\tau_{k}^{-})+[0-u_{n}\left(  \tau_{k}^{+}\right)
]\cdot1=-u_{n}\left(  \tau_{k}^{+}\right)  ,\text{ for }j\leq\xi
\label{eq:wSpJumpW3}%
\end{equation}
\newline Combining the above results, we have for \emph{Case 3.2}:%
\begin{equation}
\frac{\partial s_{n}}{\partial w_{n,j}}(\tau_{k}^{+})=\left\{
\begin{array}
[c]{cl}%
0, & \text{if }u_{n}\left(  \tau_{k}^{-}\right)  =\pm1,\text{ }u_{n}\left(
\tau_{k}^{+}\right)  =0\\
\mp1, & \text{if }u_{n}\left(  \tau_{k}^{-}\right)  =0,\text{ }u_{n}\left(
\tau_{k}^{+}\right)  =\pm1
\end{array}
\right.  \label{eq:wSpJumpWs}%
\end{equation}
Finally, note that $\frac{\partial s_{n}}{\partial w_{n,\xi}}(t)=0$ for
$t\in\lbrack0,\tau_{k})$, since the position of the agent $n$ cannot be
affected by $w_{n,\xi}$ prior to such an event.

\textbf{Objective Function Gradient Evaluation.} Proceeding as in the
evaluation of $\nabla J(\theta)$ in Section \ref{sec:oneagent}, we are now
interested in minimizing the objective function $J(\theta,w)$ in
(\ref{eq:costfunction}) with respect to $\theta$ and $w$ and we can obtain
$\nabla J(\theta,w)=[\frac{dJ\left(  \theta,w\right)  }{d\theta}%
\frac{dJ\left(  \theta,w\right)  }{dw}]^{\mathtt{T}}$ as
\[
\nabla J(\theta,w)=\frac{1}{T}\sum_{i=1}^{M}\sum_{k=0}^{K}%
%TCIMACRO{\dint _{\tau_{k}(\theta,w)}^{\tau_{k+1}(\theta,w)}}%
%BeginExpansion
{\displaystyle\int_{\tau_{k}(\theta,w)}^{\tau_{k+1}(\theta,w)}}
%EndExpansion
\nabla R_{i}\left(  t\right)  dt
\]
This depends entirely on $\nabla R_{i}\left(  t\right)  $, which is obtained
from (\ref{eq:sSpt}) and (\ref{eq:wSpt1}) and the event times $\tau_{k}$,
$k=1,\ldots,K$, given initial conditions $s_{n}\left(  0\right)  =a$ for
$n=1,\ldots,N$, and $R_{i}\left(  0\right)  $ for $i=1,\ldots,M$. In
(\ref{eq:sSpt}), $\frac{\partial R_{i}}{\partial\theta_{n,\xi}}\left(
\tau_{k}^{+}\right)  $ is obtained through (\ref{eq:wRi}) and (\ref{eq:wRpR0}%
), whereas $\frac{\partial s_{n}\left(  \tau_{k}^{+}\right)  }{\partial
\theta_{n,\xi}}$ is obtained through (\ref{eq:wSpt}), (\ref{eq:wsn}),
(\ref{eq:wSpJump}), and (\ref{eq:wSpZeroToOne2}). In (\ref{eq:wSpt1}),
$\frac{\partial R_{i}}{\partial w_{n,\xi}}\left(  \tau_{k}^{+}\right)  $ is
again obtained through (\ref{eq:wRi}) and (\ref{eq:wRpR0}), whereas
$\frac{\partial s_{n}\left(  \tau_{k}^{+}\right)  }{\partial w_{n,\xi}}$ is
obtained through (\ref{eq:wsn}), and (\ref{eq:wSpJumpWs}).

\textbf{Remark 2}. Observe that the evaluation of $\nabla R_{i}\left(
t\right)  $, hence $\nabla J(\theta,w)$, is independent of $A_{i}$,
$i=1,\ldots,M$, i.e., the values in our uncertainty model. In fact, the
dependence of $\nabla R_{i}\left(  t\right)  $ on $A_{i}$, $i=1,\ldots,M$,
manifests itself through the event times $\tau_{k}$, $k=1,\ldots,K$, that do
affect this evaluation, but they, unlike $A_{i}$ which may be unknown, are
directly observable during the gradient evaluation process. Thus, the IPA
approach possesses an inherent \emph{robustness} property: there is no need to
explicitly model how uncertainty affects $R_{i}(t)$ in (\ref{eq:multiDynR}).
Consequently, we may treat $A_{i}$ as unknown without affecting the solution
approach (the values of $\nabla R_{i}\left(  t\right)  $ are obviously
affected). We may also allow this uncertainty to be modeled through random
processes $\{A_{i}(t)\}$, $i=1,\ldots,M$; in this case, however, the result of
Proposition \ref{lem:fullSpeedStop} no longer applies without some conditions
on the statistical characteristics of $\{A_{i}(t)\}$ and the resulting $\nabla
J(\theta,w)$ is an estimate of a stochastic gradient.)

\subsection{\textbf{Objective Function Optimization}}

We now seek to obtain $\theta^{\star}$ and $w^{\ast}$ minimizing $J(\theta,w)$
through a standard gradient-based optimization scheme of the form%
\begin{equation}
\lbrack\theta^{l+1}w^{l+1}]^{\mathtt{T}}=[\theta^{l}w^{l}]^{\mathtt{T}%
}-\left[  \eta_{\theta}\text{ }\eta_{w}\right]  \tilde{\nabla}J(\theta
^{l},w^{l}) \label{eq:updatetheta}%
\end{equation}
where $\{\eta_{\theta}^{l}\},\{\eta_{w}^{l}\}$ are appropriate step size
sequences and $\tilde{\nabla}J(\theta^{l},w^{l})$ is the projection of the
gradient $\nabla J(\theta^{l},w^{l})$ onto the feasible set (the set of
$\theta^{l+1}$ satisfying the constraint \eqref{eq:thetaconstraint}, $a\leq$
$\theta^{l+1}\leq b$, and $w^{l}\geq0$). The optimization scheme terminates
when $|\tilde{\nabla}J(\theta,w)|<\varepsilon$ (for a fixed threshold
$\varepsilon$) for some $\theta$ and $w$. Our IPA-based algorithm to obtain
$\theta^{\star}$ and $w^{\ast}$ minimizing $J(\theta,w)$ is summarized in
Algorithm \ref{alg:IPA} where we have adopted the Armijo method in step-size
selection (see \cite{polak1997optimization}) for $\{\left[  \eta_{\theta}%
^{l}\text{ }\eta_{w}^{l}\right]  \}$.

One of the unusual features in (\ref{eq:updatetheta}) is the fact that the
dimension $\Gamma_{n}^{\ast}$ of $\theta_{n}^{\star}$ and $w_{n}^{\ast}$ is a
priori unknown (it depends on $T$). Thus, the algorithm must implicitly
determine this value along with $\theta_{n}^{\star}$ and $w_{n}^{\ast}$. One
can search over feasible values of $\Gamma_{n}\in\{1,2,\ldots\}$ by starting
either with a lower bound $\Gamma_{n}=1$ or an upper bound to be found. The
latter approach results in much faster execution and is followed in Algorithm
\ref{alg:IPA}. An upper bound is determined by observing that $\theta_{n,\xi}$
is the switching point where agent $n$ changes speed from $1$ to $0$ for $\xi$
odd and from $-1$ to $0$ for $\xi$ even. By setting these two groups of
switching points so that their distance is sufficiently small and waiting
times $w_{n}=\mathbf{0}$ for each agent, we determine an approximate upper
bound for $\Gamma_{n}$ as follows. First, we divide the feasible space $[a,b]$
evenly into $N$ intervals: $[a+\frac{n-1}{N}\left(  b-a\right)  ,a+\frac{n}%
{N}\left(  b-a\right)  ]$, $n=1,\ldots,N$. Define $D_{n}=a+\frac{2n-1}%
{2N}\left(  b-a\right)  $ to be the geometric center of each interval and set
\begin{equation}
\left\{
\begin{array}
[c]{cc}%
\theta_{n,\xi}=D_{n}-\sigma & \text{if }\xi\text{ even}\\
\theta_{n,\xi}=D_{n}+\sigma & \text{if }\xi\text{ odd}%
\end{array}
\right.  \label{eq:InitSwitchP}%
\end{equation}
so that the distance between switching points $\theta_{n,\xi}$ for $\xi$ odd
and even is $2\sigma$, where $\sigma>0$ is an arbitrarily small number,
$n=1,\ldots,N.$ In addition, set $w_{n}=\mathbf{0}$. Then, $T$ must satisfy%

\begin{equation}
\theta_{n,1}-s_{n}\left(  0\right)  +2\sigma\left(  \Gamma_{n}-1\right)  \leq
T\leq\theta_{n,1}-s_{n}\left(  0\right)  +2\sigma\Gamma_{n}
\label{eq:InitNumOfSwitch}%
\end{equation}
$n=$ $1,\ldots,N$, where $\Gamma_{n}$ is the number of switching points agent
$n$ can reach during $(0,T]$, given $\theta_{n,\xi}$ are defined in
(\ref{eq:InitSwitchP}). From (\ref{eq:InitNumOfSwitch}) and noting that
$\Gamma_{n}$ is an integer, we have%
\begin{equation}
\Gamma_{n}=\left\lceil \frac{1}{2\sigma}\left[  T-\theta_{n,1}+s_{n}\left(
0\right)  \right]  \right\rceil \label{eq:GammaFunc}%
\end{equation}
where $\left\lceil \cdot\right\rceil $ is the ceiling function. Clearly,
reducing $\sigma$ increases the initial number of switching points $\Gamma
_{n}$ assigned to agent $n$ and $\Gamma_{n}\rightarrow\infty$ as
$\sigma\rightarrow0$. Therefore, $\sigma$ is selected sufficiently small while
ensuring that the algorithm can be executed sufficiently fast.

As Algorithm \ref{alg:IPA} repeats steps 3-6, $w_{n,\xi}\geq0$ and distances
between $\theta_{n,\xi}$ for $\xi$ odd and even generally increase, so that
the number of switching points agent $n$ can actually reach within $T$
decreases. In other words, as long as $\sigma$ is sufficiently small (hence,
$\Gamma_{n}$ is sufficiently large), when the algorithm converges to a local
minimum and stops, there exists $\zeta_{n}<\Gamma_{n}$, such that
$\theta_{n,\zeta_{n}}$ is the last switching point agent $n$ can reach within
$(0,T]$, $n=1,\ldots,N.$ Observe that there generally exist $\xi$ such that
$\zeta_{n}<\xi\leq$ $\Gamma_{n}$ which correspond to points $\theta_{n,\xi}$
that agent $n$ cannot reach within $(0,T]$; the associated derivatives of the
cost with respect to such $\theta_{n,\xi}$ are $0$, since perturbations to
these $\theta_{n,\xi}$ will not affect $s_{n}\left(  t\right)  $, $t\in(0,T]$
and thus the cost $J(\theta,w).$ When $|\tilde{\nabla}J(\theta,w)|<\epsilon$,
we achieve a local minimum and stop, at which point the dimension of
$\theta_{n}^{\star}$ and $w_{n}^{\ast}$ is $\zeta_{n}$.

\begin{algorithm}
\caption{: IPA-based optimization algorithm to find $\theta^{\star}$ and $w^{\star}$}
\label{alg:IPA}
\begin{algorithmic}[1]
\STATE Pick $\sigma>0$ and $\epsilon>0$.
\STATE Define $D_{n}=a+\frac
{2n-1}{2N}\left(  b-a\right)  ,n$ $=1,\ldots,N,$ and set $\left\{
\begin{array}
[c]{cc}%
\theta_{n,\xi}=D_{n}-\sigma & \text{if }\xi\text{ even}\\
\theta_{n,\xi}=D_{n}+\sigma & \text{if }\xi\text{ odd}%
\end{array}
\right.$. \\Set
$w=[w_{1},\ldots,w_{N}]=0.$, where $w_{n}=[w_{n,1},\ldots,w_{n,\xi_{n}}]$ and $\Gamma_{n}=\left\lceil \frac{1}{2\sigma}\left[  T-\theta_{n,1}+s_{n}\left(
0\right)  \right]  \right\rceil$
\REPEAT
\STATE Compute $s_{n}(t)$,  $t\in[0,T]$ using $s_{n}(0)$, (\ref{eq:multiUstar}), $\theta$ and $w$ for $n=1,\ldots,N$
\STATE Compute $\tilde\nabla J(\theta,w)$ and update $\theta,w$ through \eqref{eq:updatetheta}
\UNTIL{$|\tilde\nabla J(\theta,w)|<\epsilon$}
\STATE Set $\theta_{n}^{\star} = \left[  \theta_{n,1}^{\ast},\ldots,\theta_{n,\zeta_{n}}^{\ast}\right]$ and $w_{n}^{\star} = \left[  w_{n,1}^{\ast},\ldots,w_{n,\zeta_{n}}^{\ast}\right]$, where $\zeta_{n}$ is the index of $\theta_{n,\zeta_{n}}$, which is the
last switching point agent $n$ can reach within $(0,T]$, $n=1,\ldots,N$
\end{algorithmic}
\end{algorithm}

\section{Numerical Results}

\label{sec:exper}

\begin{figure}[ptb]
\centering
\subfigure[$a=0,b=20$. $A_{i}=0.1,i=1,\ldots ,20$. $J^{\ast}=17.77.$]{
\label{fig:OneAgentTraCost1}
\includegraphics[
height=2.26in,
width=3in]
{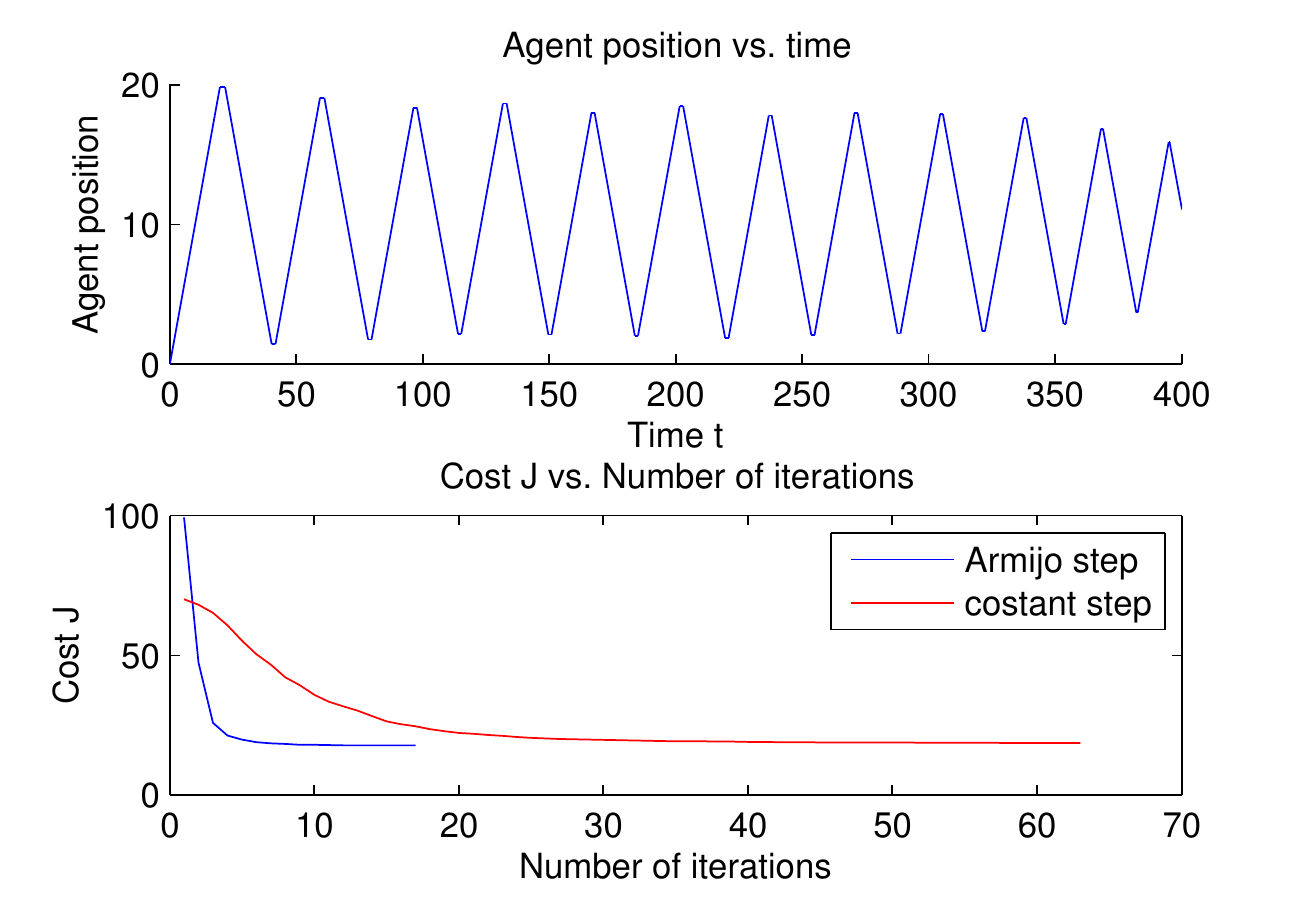}}
\subfigure[$a=4,b=16$. $A_{i}=0.1,i=1,\ldots ,20$. $J^{\ast}=39.14.$]{
\label{fig.OneAgentTraCost2}
\includegraphics[
height=2.26in,
width=3in]
{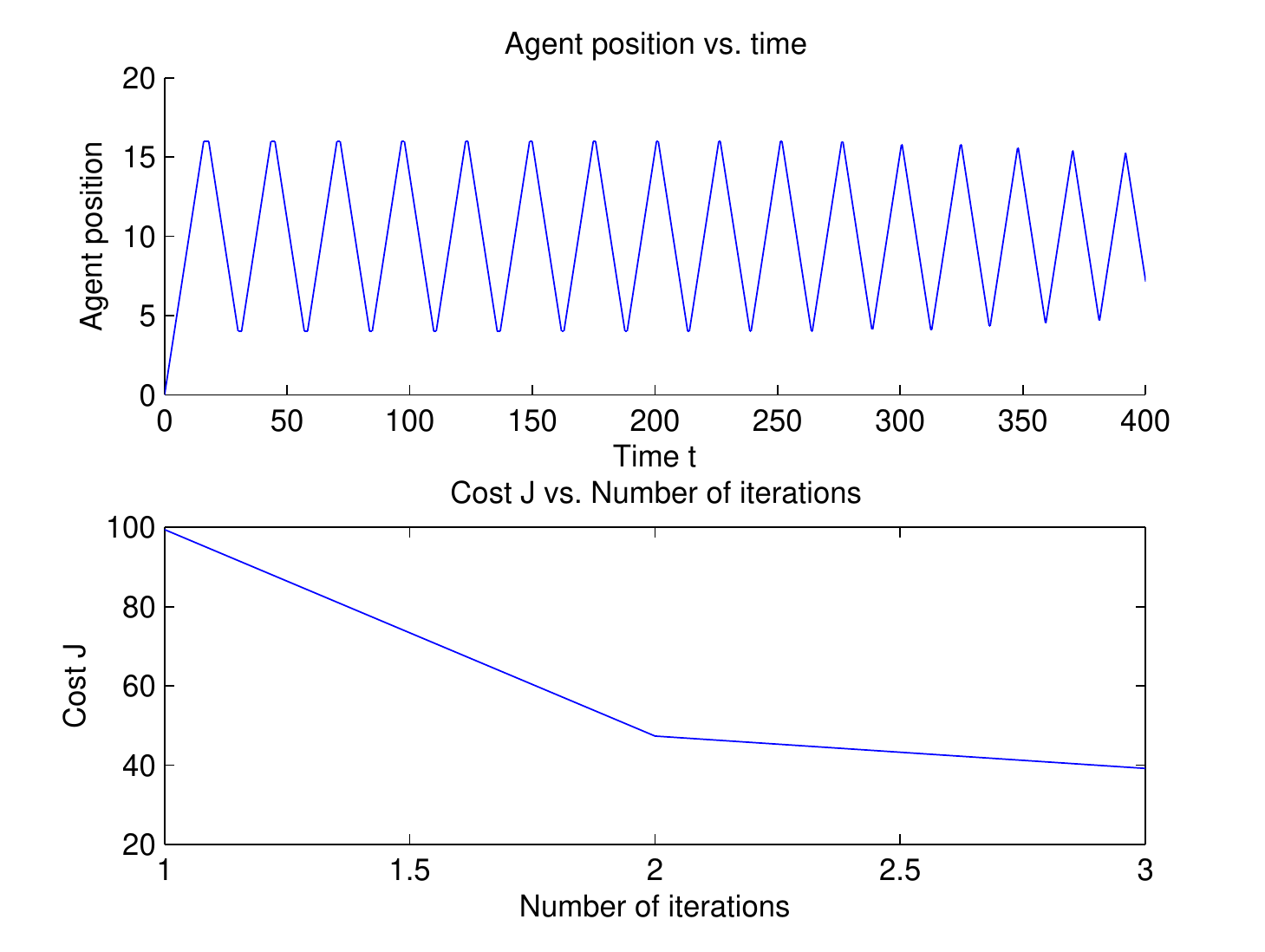}}
\subfigure[$a=0,b=20$. $A_{0}=A_{20}=0.5,A_{i}=0.1,i=1,\ldots ,19.$ $J^{\ast}=39.30.$]{
\label{fig:OneAgentTraCost1}
\includegraphics[
height=2.26in,
width=3in]
{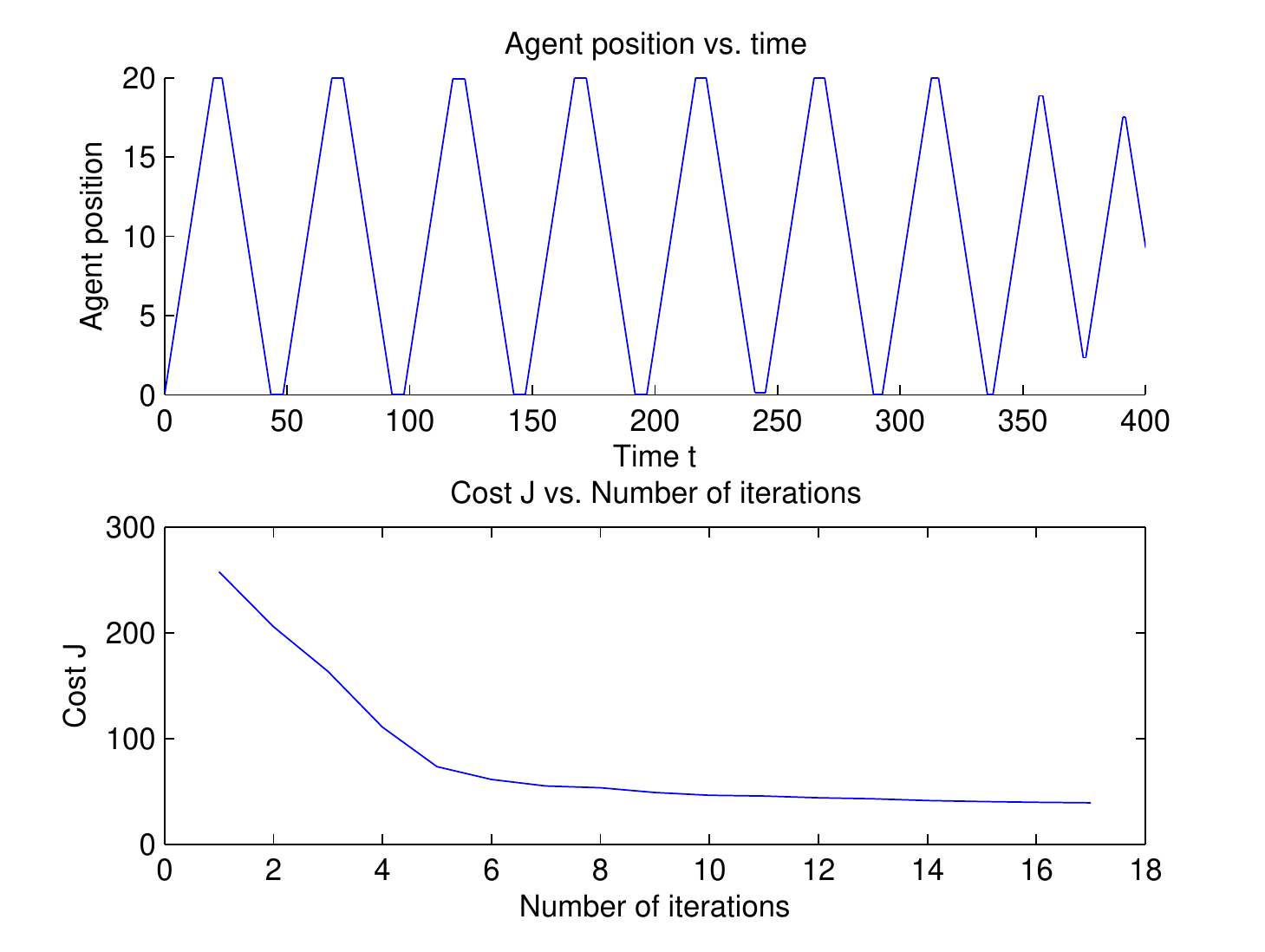}}
\subfigure[$a=0,b=20$. $A_{i}\left( \Delta t_{i}\right) \backsim U\left( 0.075,0.125\right) $, $\Delta t_{i}\backsim 0.1e^{-0.1t}$. $J^{\ast}=17.54.$]{
\label{fig:OneAgentTraCost1}
\includegraphics[
height=2.26in,
width=3in]
{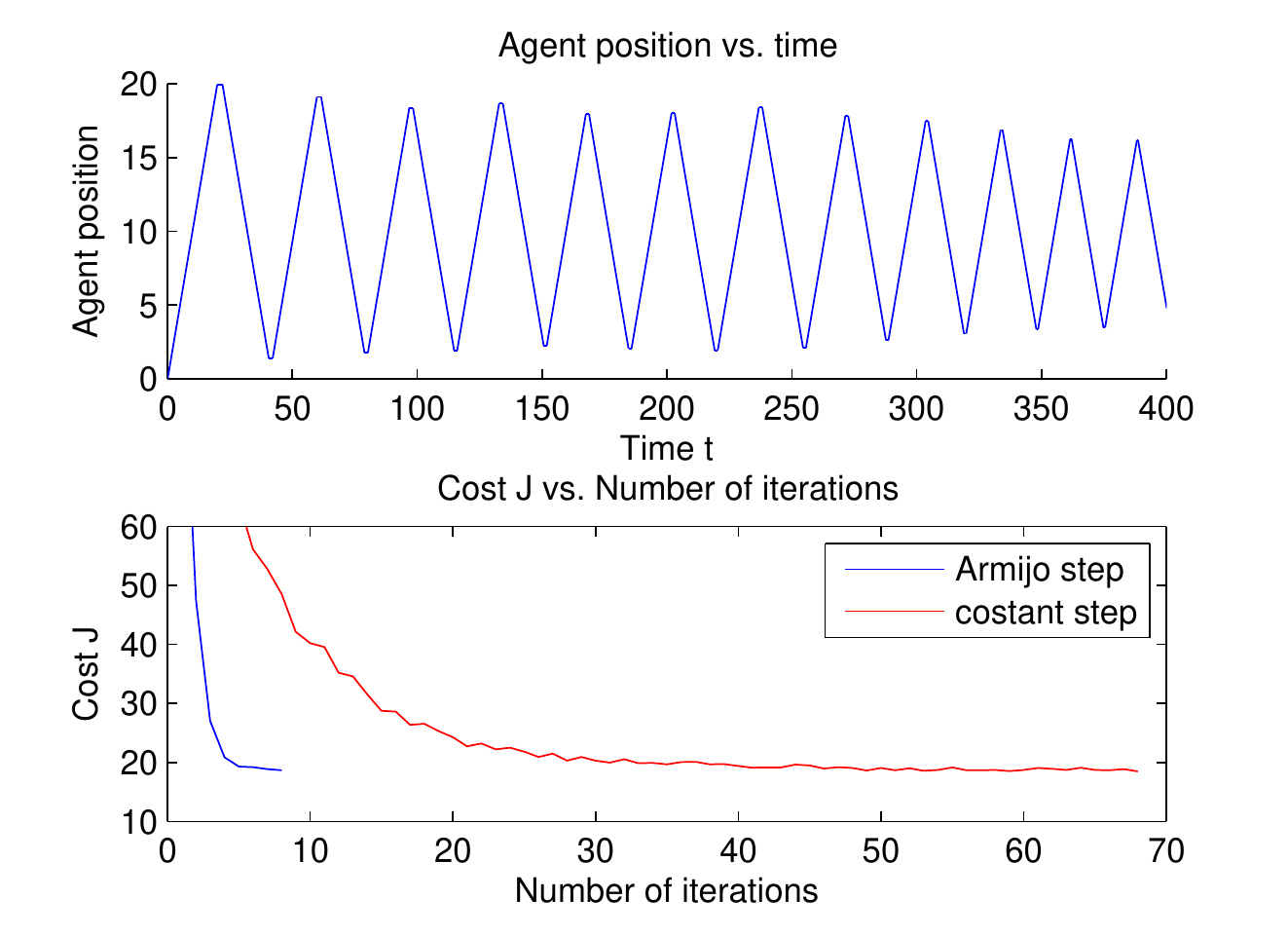}}\caption{One agent example. $L=20,T=400$. For each
example, top plot: optimal trajectory; bottom plot: J versus iterations.}%
\label{fig:OneAgentTraCost}%
\end{figure}

\begin{figure}[ptb]
\centering
\subfigure[$a=0,b=20$.]{
\label{fig:OneAgentMagTra1}
\includegraphics[
height=1.2in,
width=3in]
{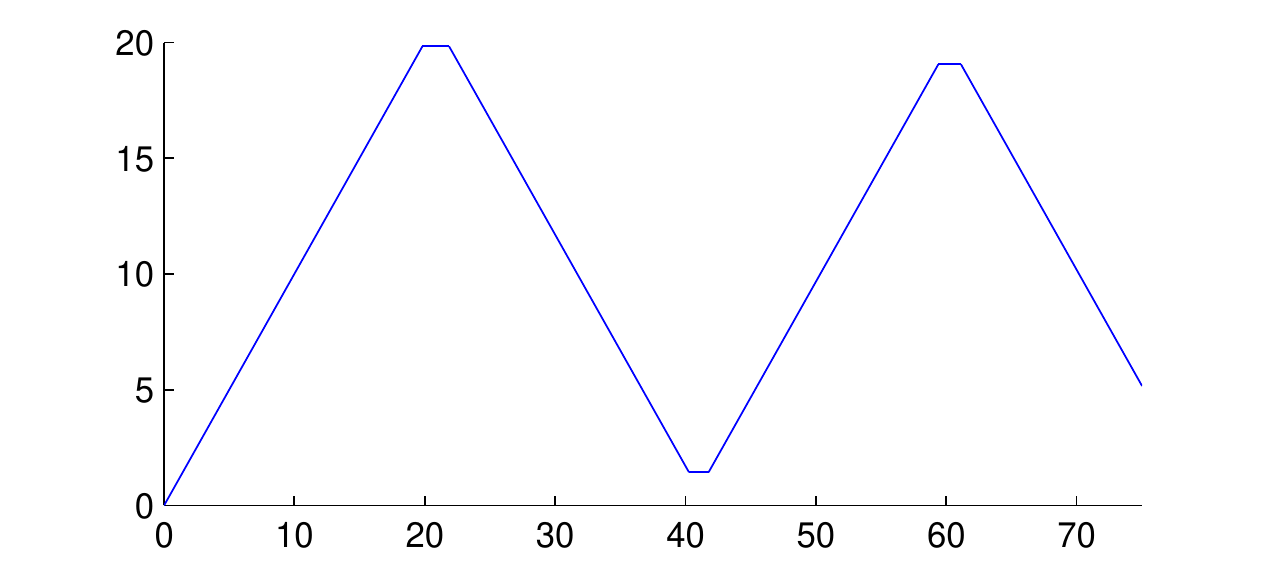}} \subfigure[$a=4,b=16$.]{
\label{fig:OneAgentMagTra2}
\includegraphics[
height=1.2in,
width=3in]
{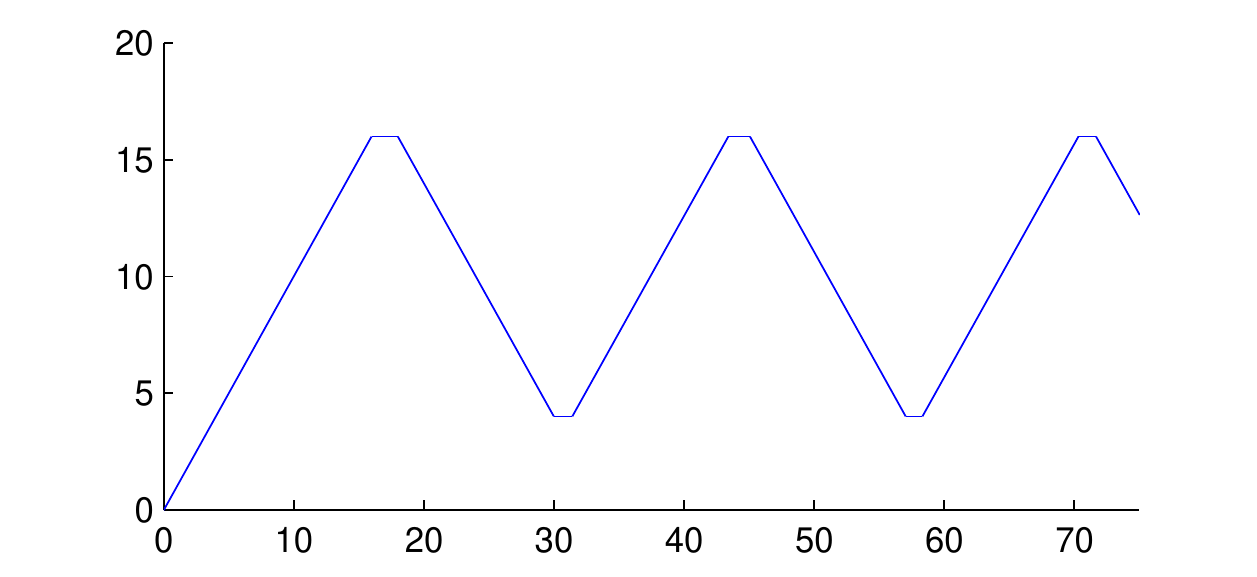}}\caption{Magnified trajectory for sub-figure (a) and (b) in
Fig. \ref{fig:OneAgentTraCost}, $t\in\left[  0,75\right]  $.}%
\label{fig:OneAgentMagTra}%
\end{figure}

\begin{figure}[ptb]
\centering
\subfigure[$a=0,b=20$.]{
\label{fig:OneAgentTable1}
\includegraphics[
height=2.98in,
width=1.9in]
{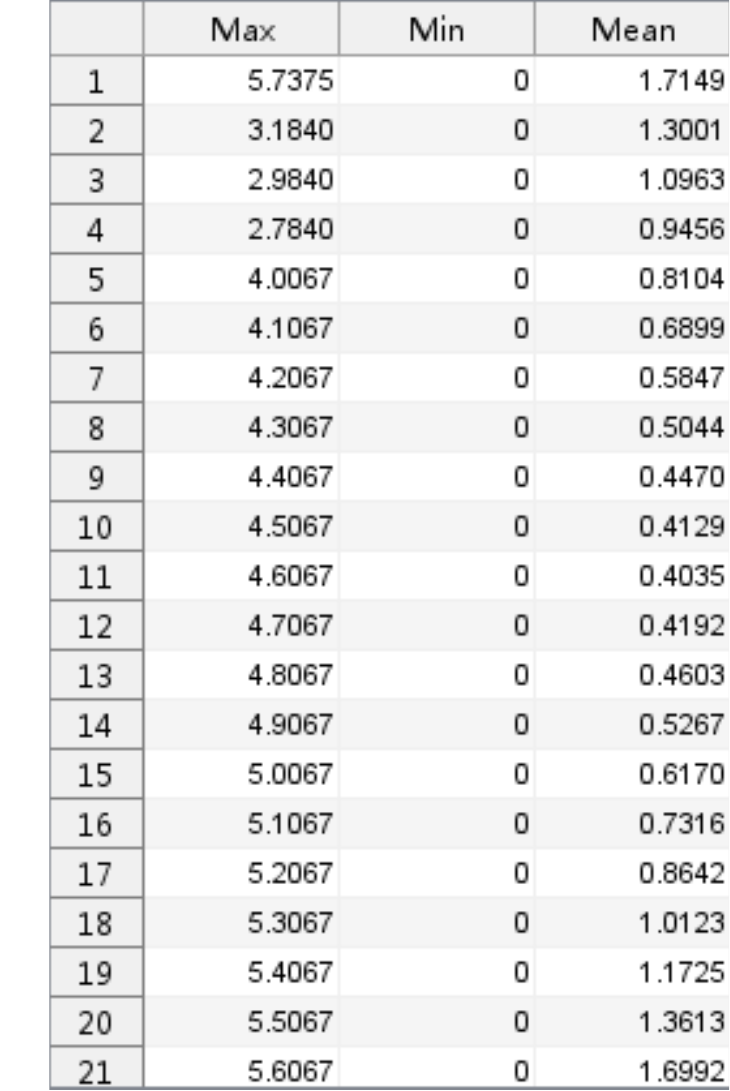}} \subfigure[$a=4,b=16$.]{
\label{fig:OneAgentTable2}
\includegraphics[
height=2.98in,
width=1.9in]
{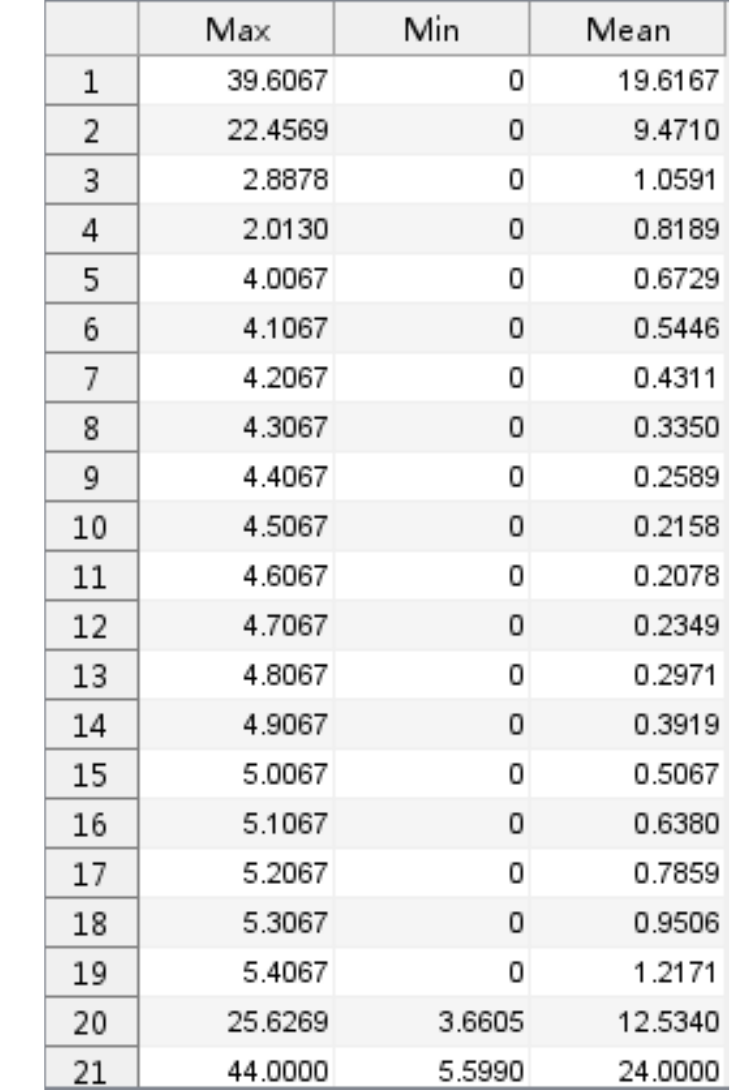}}\caption{Max, min and mean uncertainty value for each sampling
point.}%
\label{fig:OneAgentTable}%
\end{figure}

In this section we present some examples of persistent monitoring problems in
which agent trajectories are determined using Algorithm \ref{alg:IPA}. The
first four are one-agent examples with $L=20$, $M=21$, $\alpha_{1}=0$,
$\alpha_{M}=20$, and the remaining sampling points are evenly spaced over
$[0,20]$. The sensing range in (\ref{eq:multiLinearModel}) is set to $r=4$,
the initial values of the uncertainty functions in (\ref{eq:multiDynR}) are
$R_{i}(0)=4$ for all $i$, and the time horizon is $T=400$. In Fig.
\ref{fig:OneAgentTraCost}(a) we show results where the agent is allowed to
move over the entire space $[0,20]$ and the uncertainty model is selected so
that $B=3$ and $A_{i}=0.1$ for all $i=1,\ldots,20$, whereas in Fig.
\ref{fig:OneAgentTraCost}(b) the feasible space is limited to $[a,b]$ with
$a=r=4$ and $b=L-r=16$. The top plot in each example shows the optimal
trajectory $s^{\ast}(t)$ obtained, while the bottom shows the cost
$J(\theta^{l},w^{l})$ as a function of iteration number. In Fig.
\ref{fig:OneAgentMagTra}, the trajectories in Fig. \ref{fig:OneAgentTraCost}%
(a),(b) are magnified for the interval $t\in\lbrack0,75]$ to emphasize the
presence of strictly positive waiting times at the switching points. In
addition, maximum, minimum and mean values for the uncertainty function of
each sampling point in these two cases are shown in Fig.
\ref{fig:OneAgentTable}. Observe that when $a=r$ and $b=L-r$, an instability
arises at the last two sampling points of both ends of the mission space; this
is expected since the agent's sensing range can only marginally reach the two
end points from $a=4$ and $b=16$.

In Fig. \ref{fig:OneAgentTraCost}(c) we show results for a case similar to
Fig. \ref{fig:OneAgentTraCost}(a) except that the values of $A_{i}$ are
selected so that $A_{0}=A_{20}=0.5$, while $A_{i}=0.1,i=1,\ldots,19$. We
should point out that even though it seems that the trajectory includes
switching points at the two end points, this is not the case: the switching
points are very close but not equal to these end points, consistent with
Proposition \ref{lem:switchingpoints}. In Fig. \ref{fig:OneAgentTraCost}(d),
on the other hand, the values of $A_{i}$ are allowed to be random, thus
dealing with a persistent monitoring problem in a stochastic mission space. In
particular, each $A_{i}$ is treated as a piecewise constant random process
$\{A_{i}(t)\}$ such that $A_{i}(t)$ takes on a fixed value uniformly
distributed over $\left(  0.075,0.125\right)  $ for an exponentially
distributed time interval with mean $10$ before switching to a new value. Note
that the behavior of the system in this case is very similar to Fig.
\ref{fig:OneAgentTraCost}(a) where $A_{i}=0.1$ for all $i=1,\ldots,20$ without
any change in the way in which $\nabla J(\theta^{l},w^{l})$ is evaluated in
executing (\ref{eq:updatetheta}). As already pointed out, this exploits a
robustness property of IPA which makes it independent of the values of $A_{i}%
$. In general, however, when $A_{i}(t)$ is not time-invariant, Proposition
\ref{lem:fullSpeedStop} may no longer apply, since an extra term $\sum_{i}%
\dot{A}_{i}\left(  t\right)  $ would be present in (\ref{eq:HtimeInvar3}). In
such a case, $u_{n}^{\ast}\left(  t\right)  $ may be nonzero when $\lambda
_{n}^{\ast}\left(  t\right)  =0$ and the determination of an optimal
trajectory through switching points and waiting times alone may no longer be
possible. In the case of \ref{fig:OneAgentTraCost}(d), $A_{i}(t)$ changes
sufficiently slowly to maintain the validity of Proposition
\ref{lem:fullSpeedStop} over relatively long time intervals, under the
assumption that w.p. 1 no event time coincides with the jump times in any
$\{A_{i}(t)\}$.

In all cases, we initialize the algorithm with $\sigma=5$ and $\varepsilon
=2\times10^{-10}$. The algorithm running times are approximately 10 sec using
Armijo step-sizes. Note that although the number of iterations for the
examples shown may substantially vary, the actual algorithm running times do
not. This is simply because the Armijo step-size method may involve several
trials per iteration to adjust the step-size in order to achieve an adequate
decrease in cost. In Fig. \ref{fig:OneAgentTraCost}(a),(d), red line shows $J$
vs. number of iterations using constant step size and they almost converges to
the same optimal value. Non-smoothness in Fig. \ref{fig:OneAgentTraCost}(d)
comes from the fact that it is a stochastic system. Note that in all cases the
initial cost is significantly reduced indicating the importance of optimally
selecting the values of the switching points and associated waiting times (if any).

\begin{figure}[ptb]
\centering
\subfigure[$a=0,b=20$. $J^{\ast}=17.77.$]{
\label{fig:TwoAgentTraCost1}
\includegraphics[
height=2.26in,
width=3in]
{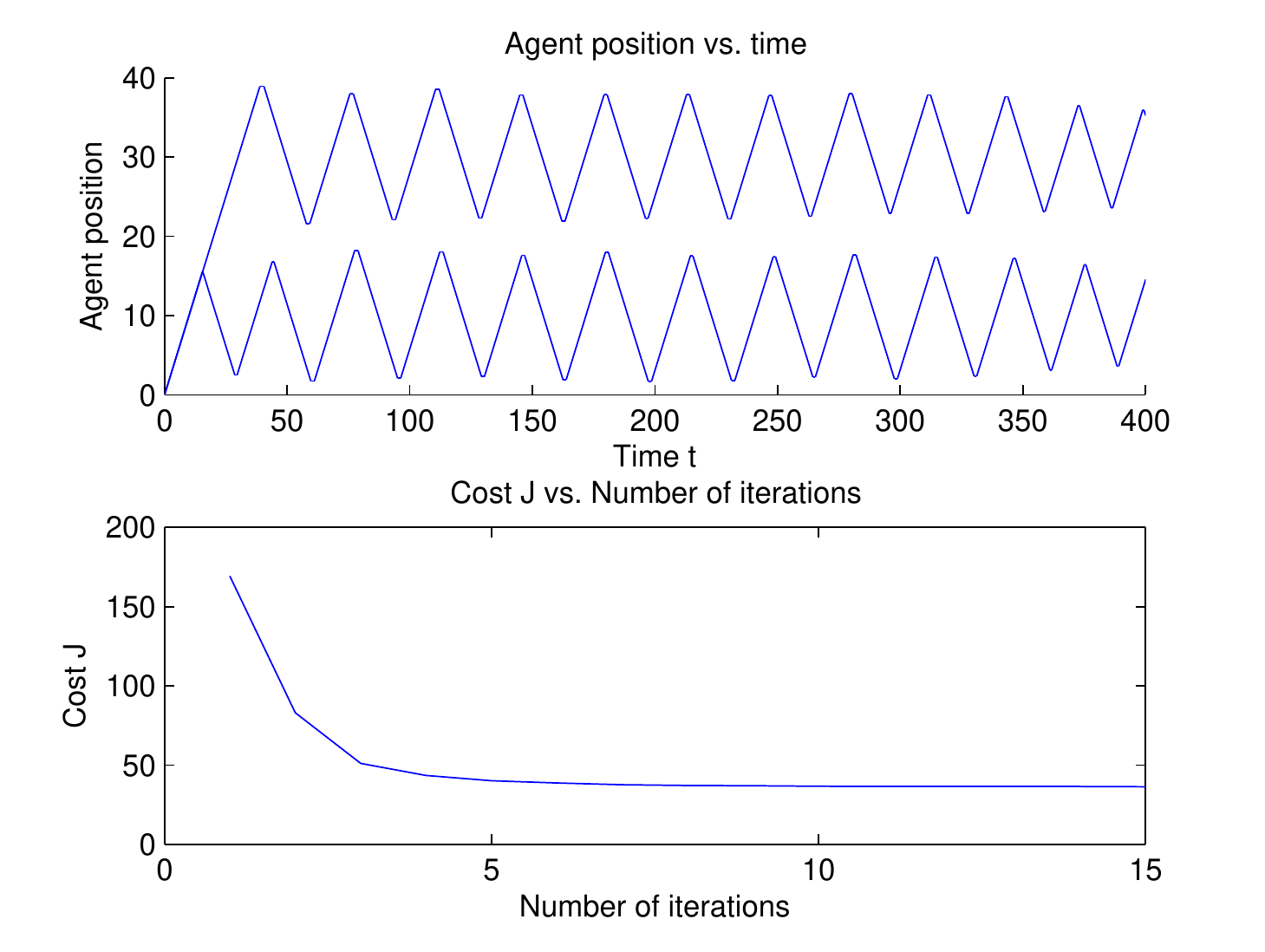}} \subfigure[$a=4,b=16$. $J^{\ast}=39.14.$]{
\label{fig:TwoAgentTraCost2}
\includegraphics[
height=2.26in,
width=3in]
{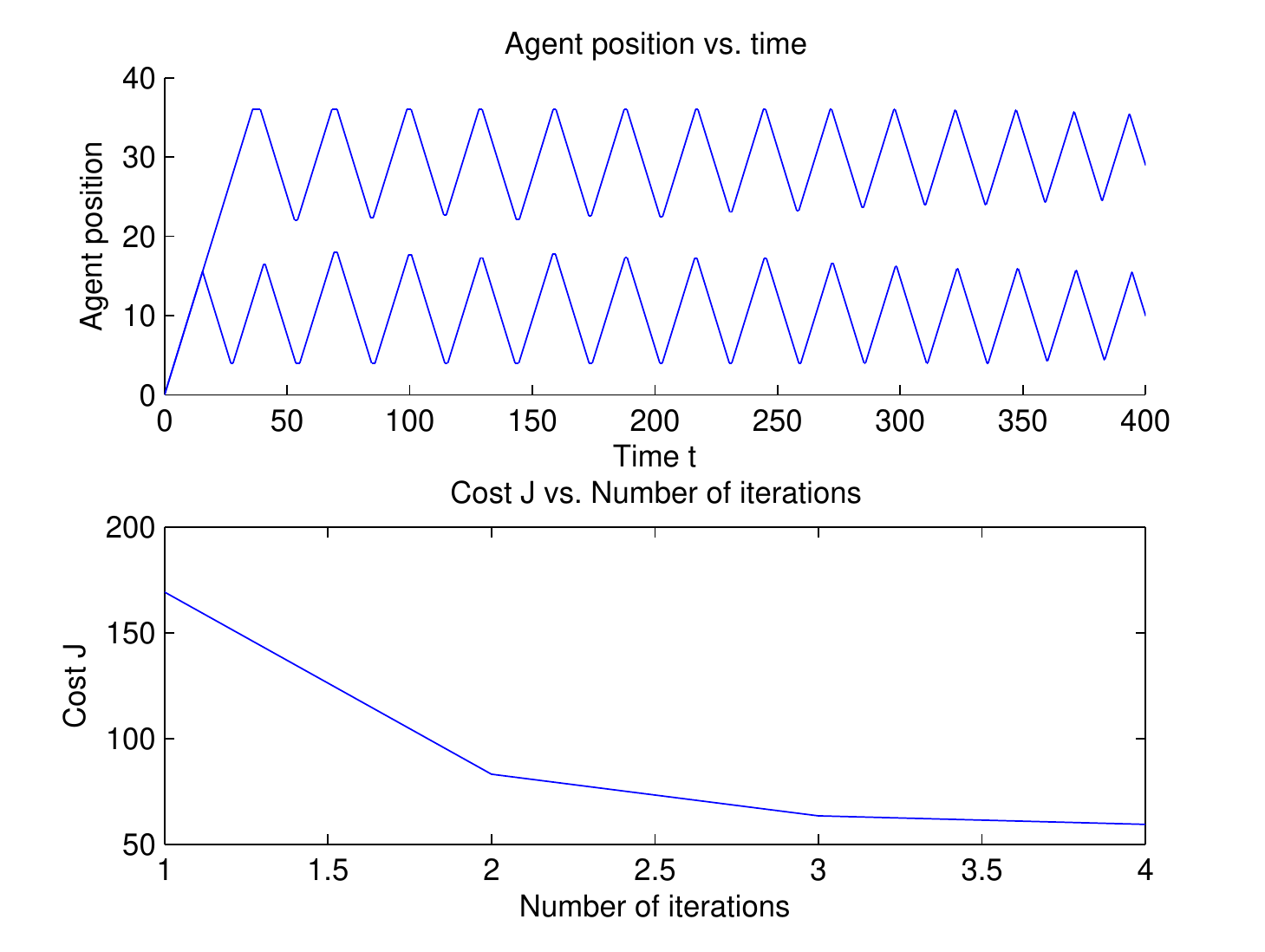}}\caption{Two agent example. $L=40,T=400$. Top plot: optimal
trajectory. Bottom plot: J versus iterations.}%
\label{fig:TwoAgentTraCost}%
\end{figure}

Figure \ref{fig:TwoAgentTraCost} shows two two-agent examples with $L=40$,
$M=41$ and evenly spaced sampling points over $[0,L]$, $A_{i}=0.01,$ $B=3,$
$r=4,$ $R_{i}(0)=4$ for all $i$ and $T=400$. In Fig. \ref{fig:TwoAgentTraCost}%
(a) the agents are allowed to move over the whole mission space $[0,L]$, while
in Fig. \ref{fig:TwoAgentTraCost}(b) they are only allowed to move over
$[a,b]$ where $a=r$ and $b=L-r$. We initialize the algorithm with the same
$\sigma$ and $\varepsilon$ as before. The algorithm running time is
approximately 15 sec using Armijo step-sizes, and we observe once again
significant reductions in cost.

\section{Conclusion}

\label{sec:concl}

\bibliographystyle{IEEEtran}
\bibliography{Papers}

% Generated by IEEEtran.bst, version: 1.13 (2008/09/30)
\begin{thebibliography}{10}
\providecommand{\url}[1]{#1}
\csname url@samestyle\endcsname
\providecommand{\newblock}{\relax}
\providecommand{\bibinfo}[2]{#2}
\providecommand{\BIBentrySTDinterwordspacing}{\spaceskip=0pt\relax}
\providecommand{\BIBentryALTinterwordstretchfactor}{4}
\providecommand{\BIBentryALTinterwordspacing}{\spaceskip=\fontdimen2\font plus
\BIBentryALTinterwordstretchfactor\fontdimen3\font minus
  \fontdimen4\font\relax}
\providecommand{\BIBforeignlanguage}[2]{{%
\expandafter\ifx\csname l@#1\endcsname\relax
\typeout{** WARNING: IEEEtran.bst: No hyphenation pattern has been}%
\typeout{** loaded for the language `#1'. Using the pattern for}%
\typeout{** the default language instead.}%
\else
\language=\csname l@#1\endcsname
\fi
#2}}
\providecommand{\BIBdecl}{\relax}
\BIBdecl

\bibitem{rekleitis2004limited}
I.~Rekleitis, V.~Lee-Shue, A.~New, and H.~Choset, ``Limited communication,
  multi-robot team based coverage,'' in \emph{Robotics and Automation, 2004.
  Proceedings. ICRA'04. 2004 IEEE International Conference on}, vol.~4.\hskip
  1em plus 0.5em minus 0.4em\relax IEEE, 2004, pp. 3462--3468.

\bibitem{cortes2004coverage}
J.~Cortes, S.~Martinez, T.~Karatas, and F.~Bullo, ``Coverage control for mobile
  sensing networks,'' \emph{Robotics and Automation, IEEE Transactions on},
  vol.~20, no.~2, pp. 243--255, 2004.

\bibitem{li2006cooperative}
W.~Li and C.~Cassandras, ``A cooperative receding horizon controller for
  multivehicle uncertain environments,'' \emph{IEEE Transactions on Automatic
  Control}, vol.~51, no.~2, pp. 242--257, 2006.

\bibitem{girard2005border}
A.~Girard, A.~Howell, and J.~Hedrick, ``Border patrol and surveillance missions
  using multiple unmanned air vehicles,'' in \emph{43rd IEEE Conference on
  Decision and Control}, vol.~1.\hskip 1em plus 0.5em minus 0.4em\relax IEEE,
  2005, pp. 620--625.

\bibitem{grocholsky2006cooperative}
B.~Grocholsky, J.~Keller, V.~Kumar, and G.~Pappas, ``Cooperative air and ground
  surveillance,'' \emph{IEEE Robotics \& Automation Magazine}, vol.~13, no.~3,
  pp. 16--25, 2006.

\bibitem{smith2012persistent}
S.~Smith, M.~Schwager, and D.~Rus, ``Persistent robotic tasks: Monitoring and
  sweeping in changing enviroments,'' \emph{IEEE Transactions on Robotics},
  2012, to appear.

\bibitem{paley2008cooperative}
D.~Paley, F.~Zhang, and N.~Leonard, ``Cooperative control for ocean sampling:
  The glider coordinated control system,'' \emph{IEEE Transactions on Control
  Systems Technology}, vol.~16, no.~4, pp. 735--744, 2008.

\bibitem{bertsimas1993stochastic}
D.~Bertsimas and G.~Van~Ryzin, ``Stochastic and dynamic vehicle routing in the
  {E}uclidean plane with multiple capacitated vehicles,'' \emph{Operations
  Research}, pp. 60--76, 1993.

\bibitem{cooper1981introduction}
R.~Cooper, \emph{Introduction to queuing theory}.\hskip 1em plus 0.5em minus
  0.4em\relax Edward Arnold, 1981.

\bibitem{sun2004perturbation}
G.~Sun, C.~Cassandras, Y.~Wardi, C.~Panayiotou, and G.~Riley, ``Perturbation
  analysis and optimization of stochastic flow networks,'' \emph{Automatic
  Control, IEEE Transactions on}, vol.~49, no.~12, pp. 2143--2159, 2004.

\bibitem{nigam2008persistent}
N.~Nigam and I.~Kroo, ``Persistent surveillance using multiple unmanned air
  vehicles,'' in \emph{IEEE Aerospace Conference}.\hskip 1em plus 0.5em minus
  0.4em\relax IEEE, 2008, pp. 1--14.

\bibitem{hokayem2008persistent}
P.~Hokayem, D.~Stipanovic, and M.~Spong, ``On persistent coverage control,'' in
  \emph{Decision and Control, 2007 46th IEEE Conference on}.\hskip 1em plus
  0.5em minus 0.4em\relax IEEE, 2008, pp. 6130--6135.

\bibitem{elmaliach2008realistic}
Y.~Elmaliach, A.~Shiloni, and G.~Kaminka, ``A realistic model of
  frequency-based multi-robot polyline patrolling,'' in \emph{Proceedings of
  the 7th international joint conference on Autonomous agents and multiagent
  systems-Volume 1}.\hskip 1em plus 0.5em minus 0.4em\relax International
  Foundation for Autonomous Agents and Multiagent Systems, 2008, pp. 63--70.

\bibitem{elmaliach2007multi}
Y.~Elmaliach, N.~Agmon, and G.~Kaminka, ``Multi-robot area patrol under
  frequency constraints,'' in \emph{Robotics and Automation, 2007 IEEE
  International Conference on}.\hskip 1em plus 0.5em minus 0.4em\relax IEEE,
  2007, pp. 385--390.

\bibitem{cassandras2009perturbation}
C.~Cassandras, Y.~Wardi, C.~Panayiotou, and C.~Yao, ``Perturbation analysis and
  optimization of stochastic hybrid systems,'' \emph{European Journal of
  Control}, vol.~16, no.~6, pp. 642--664, 2010.

\bibitem{Wardietal09}
Y.~Wardi, R.~Adams, and B.~Melamed, ``A unified approach to infinitesimal
  perturbation analysis in stochastic flow models: the single-stage case,''
  \emph{IEEE Trans. on Automatic Control}, vol.~55, no.~1, pp. 89--103, 2009.

\bibitem{bryson1975applied}
A.~Bryson and Y.~Ho, \emph{Applied optimal control}.\hskip 1em plus 0.5em minus
  0.4em\relax Wiley New York, 1975.

\bibitem{egerstedt2006transition}
M.~Egerstedt, Y.~Wardi, and H.~Axelsson, ``Transition-time optimization for
  switched-mode dynamical systems,'' \emph{Automatic Control, IEEE Transactions
  on}, vol.~51, no.~1, pp. 110--115, 2006.

\bibitem{shaikh2007hybrid}
M.~Shaikh and P.~Caines, ``On the hybrid optimal control problem: Theory and
  algorithms,'' \emph{IEEE Transactions on Automatic Control}, vol.~52, no.~9,
  pp. 1587--1603, 2007.

\bibitem{xu2004optimal}
X.~Xu and P.~Antsaklis, ``Optimal control of switched systems based on
  parameterization of the switching instants,'' \emph{Automatic Control, IEEE
  Transactions on}, vol.~49, no.~1, pp. 2--16, 2004.

\bibitem{yaoc2011perturbation}
C.~Yao and C.~Cassandras, ``Perturbation analysis of stochastic hybrid systems
  and applications to resource contention games,'' \emph{Frontiers of
  Electrical and Electronic Engineeing in China}, vol. 6,3, pp. 453--467, 2011.

\bibitem{cassandras2007stochastic}
C.~Cassandras, J.~Lygeros, and C.~Press, \emph{Stochastic hybrid
  systems}.\hskip 1em plus 0.5em minus 0.4em\relax CRC/Taylor \& Francis, 2007.

\bibitem{polak1997optimization}
E.~Polak, \emph{Optimization: algorithms and consistent approximations}.\hskip
  1em plus 0.5em minus 0.4em\relax Springer Verlag, 1997.

\end{thebibliography}

\end{document}